\documentclass[aps,twocolumn]{revtex4-1}
\usepackage{amsmath,amssymb}
\usepackage{slashed}
\usepackage{graphicx}
\usepackage{bm}
\usepackage[top=1.0in,bottom=1.0in,left=1.0in,right=1.0in]{geometry}
\usepackage[colorlinks,linkcolor=blue,citecolor=blue]{hyperref}
\newcommand{\be}{\begin{equation}}
\newcommand{\ee}{\end{equation}}
\newcommand{\bea}{\begin{eqnarray}}
\newcommand{\eea}{\end{eqnarray}}
\newcommand{\ba}{\begin{eqnarray}}
\newcommand{\ea}{\end{eqnarray}}

\begin{document}

\title{Hadronic structure on the light-front  I\\
Instanton effects and quark-antiquark effective potentials}

\author{Edward Shuryak}
\email{edward.shuryak@stonybrook.edu}
\affiliation{Center for Nuclear Theory, Department of Physics and Astronomy, Stony Brook University, Stony Brook, New York 11794--3800, USA}

\author{Ismail Zahed}
\email{ismail.zahed@stonybrook.edu}
\affiliation{Center for Nuclear Theory, Department of Physics and Astronomy, Stony Brook University, Stony Brook, New York 11794--3800, USA}

\begin{abstract}
This is the first of a sequence of papers that address the non-perturbative origin of the central, 
and spin-dependent forces between quarks. Its main thrust is a focus on meson spectroscopy in the
center of mass frame. We suggest a novel ``dense instanton ensemble" model for the QCD vacuum, to
explain  the interquark forces in mesons, from quarkonia to heavy-light and light-light ones. The sequels will show how to export these interactions to the light front, and derive the corresponding
Hamiltonians, mesonic and baryonic light front wavefunctions. The ultimate aim of the series
is to bridge the gap between hadronic spectroscopy and partonic observables.
\end{abstract}

\maketitle
\section{Introduction}

The physics of hadrons is firmly based in  Quantum Chromodynamics,  a theory over half a century old.
 One might think that by now this subject has reached a solid degree of maturity with most issues settled.
 Unfortunately this is  not yet the case.
The field of non-perturbative QCD consists of several  subfields, which can be defined as follows:
\\
{\bf 1.\,}{\em hadronic spectroscopy} (masses and wave functions of mesons, baryons, pentaquarks, etc);
\\
{\bf 2.\,}{\em QCD vacuum structure} (vacuum condensates, Euclidean correlation functions, etc); 
\\
{\bf 3.\,}{\em light-front observables} (distributions amplitudes, parton distribution functions or PDFs,  etc).
\\
They are  relatively weakly connected, although  in the past decade considerable efforts to bridge
them has taken place. Our set of papers is another effort in this direction.
In this introductory section we will give a brief description of the two aforementioned subfields.
The third one will be discussed in the papers to follow.

\subsection{Hadronic structure}

This field originated in the early  1960's, and experienced a  rapid expansion in the last decade,
due to the discoveries of multi-quark hadrons in the heavy-light sector. We will not discuss these
``exotic hadrons" in this work but  focus on  the underlying physics at the origin of
the inter-quark forces. Parameterizing and understanding them is a needed step,  setting the stage for   derivation of the effective Hamiltonians
on the light front.

Early nonrelativistic quark models  achieved resounding success, in part through simple  formulae
for the masses and magnetic moments of  baryons made of light quarks. 
 The revolutionary discoveries of 1970's of ``quarkonia", the bound states made of heavy $c,b$ quarks,  revealed
  families of narrow bound states  well described
by the celebrated Cornell potential
\begin{equation}  V_{\rm Cornell}(r)=-{4 \alpha_s \over 3}{1 \over r}+\sigma_T r 
\end{equation}
Further theory effort  developed 
effective non-relativistic theories, such as non-relativistic QCD (NRQCD) 
and perturbative non-relativistic QCD (pNRQCD). They put  spectroscopy
of quarkonia close to that of atoms and nuclei in accuracy. 
A number of   $universal$ (flavor independent) central and spin-dependent potentials were defined, expressed
$via$ certain nonlocal  correlators of vacuum gauge fields.  For short but concise summary see 
e.g. \cite{Pineda:2000sz}.

The Cornell potential attributes the short-distance
potential to perturbative one-gluon exchange, and its large distance ${\cal O}(r)$ contribution  to the tension of the confining flux tube (QCD string). The issues to be discussed in this paper are the nonperturbative origins of the inter-quark interaction  at $intermediate$ distances $r\sim 0.2-0.5 \,$fm. Those are especially important for
small size-hadrons, such as bottomonia or pions (see below).

A static quark potential $V_C(r)$ is defined through a vacuum average of a pair of parallell 
Wilson lines,  
\begin{equation} e^{-V_C(r) T}=\langle W(\vec x_1) W^\dagger(\vec x_2)\rangle 
\end{equation} 
defined by
  \begin{equation} 
  W(\vec x)={\rm Pexp}\bigg(i g \int_W  dx_4 A_\mu^a(x_4,\vec x) \hat T^a \bigg)
  	\end{equation}
where color generator $\hat T^a=t^a/2$, with $t^a$ Gell-Mann matrices. $tPexp$ is path-ordered
matrix product. 
The parallel lines are running  in the Euclidean time direction, separated by the spatial distance $r=|\vec x_1 -\vec x_2|$.
The lines are connected at times $\pm T/2$ , but eventually the limit of large time $T\rightarrow \infty$ is assumed,
this is mostly unimportant.
Historically, the derivation of this potential  has been at the forefront of lattice studies 
 of the QCD vacuum. In a way, it is similar in spirit to the  Born-Oppenheimer approximation in molecular physics, 
 where the probe quarks back-reaction is ignored. For recent developments and references, we refer to
 the work by  Brambilla et al~\cite{1707.09647}.

Traditionally, the central potential was used as  linear  $V_c(r)=\sigma_T r$
with  $\sigma_T $ be the tension of the confining flux tube (known also as QCD string). 
So, one might think the issue of the central potential is very simple.
Unfortunately, it is far from being so, and there are several impediments for usage of the flux tubes.

The first  is that the simple linear term is
 expected to be correct at large distances only, while at smaller $r$ there are 
 other nonperturbative effects as well. The simplest corrections are due to
quantum vibrations of the  flux tube. The first of them is the famous   ``Lusher term",
 attractive $\sim 1/r$. If the
string is described by a Nambu-Goto geometrical action, these vibrations are summed up in the so called  Arvis form
\cite{Arvis:1983fp} 
\be  \label{eqn_Arvis}
E(r)=\sigma_T r \sqrt{ 1-{\pi \over 6} {1 \over \sigma_T r^2} } 
\ee
This expression predicts that potential vanishes at $r\approx 1/3$ fm. 
Furthermore,  the
 QCD string may have much more complicated action,
so the Arvis potential cannot be really trusted when the corrections  are large. Yet the two first expansion
terms are universal, see more in~\cite{Aharony:2010db}.  Lattice studies of 
these corrections (e.g.~\cite{Athenodorou:2021qvs}) in the setting with two static charges, or  a flux tube wrapped around
the lattice,  do confirm these two terms.

Important issue to be discussed is the derivation of the relativistic corrections $\sim (v/c)^2\sim (\Lambda_{QCD}/m_Q)^2$
where $m_Q$ is heavy quark mass and $\Lambda_{QCD}$ stands for a ``QCD scale".
Such perturbative terms are well known  in atomic and nuclear physics, and changing from one-photon to  one-gluons exchange is simple. So it is clear how {\em spin-spin, spin-orbit}
and {\em tensor force} $V_{SS},V_{SL},V_{T}$ arise from perturbative contributions, being all just certain derivatives of the Coulomb potential. 

The question is what is the nonperturbative contributions to these
potentials.
Those can be   related to  
Wilson lines  decorated by two extra field strengths \cite{Callan:1978ye,Eichten:1980mw}. Schematically, they have the form
$$ \langle W G_{\mu\nu}(x) W W^\dagger G_{\mu\nu}(y ) W^\dagger\rangle$$
More details of the invariant spin potentials given in section~\ref{sec_correlators}.

In a nonrelativistic setting, spin-dependent forces originate  from quark (gluo)-magnetic moments interacting
with vacuum fluctuations of (mostly)  {\it gluo-magnetic}
 fields. Unfortunately, the QCD flux tubes carry {\it electric} flux: models of their structure and lattice studies specify mostly distribution of 
 the gluo-electric fields in them,  not the gluo-magnetic ones
needed for the spin forces. So, one has to think about other origins of those in the vacuum.

In principle, the pertinent correlation functions can be calculated on the lattice, so one might think
that all spin-dependent forces are by now well documented and their origins explained. Unfortunately, this is not the case
as quantitative lattice studies have only started recently. 
Below  we will investigate to what extent such correlators, evaluated using certain 
models of vacuum fields, 
can reproduce the  observed spin-dependent potentials. We will find that in heavy quarkonia
 instanton-induced forces can provide contributions to perturbative ones, leading together
to a good description of splittings for a number of the low-lying states.

\begin{figure}[h!]
	\includegraphics[width=4cm]{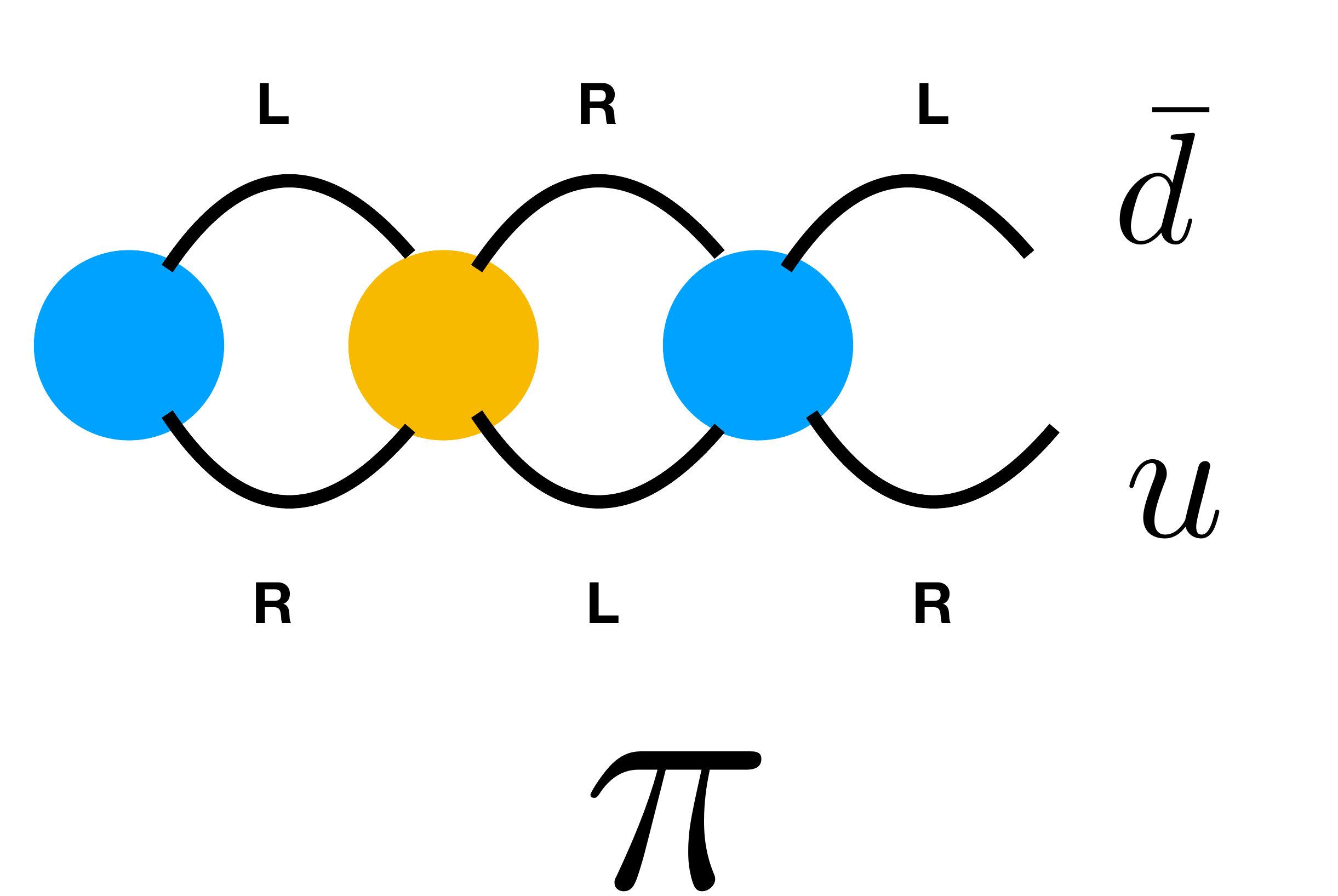}
\vskip .2cm		\includegraphics[width=4cm]{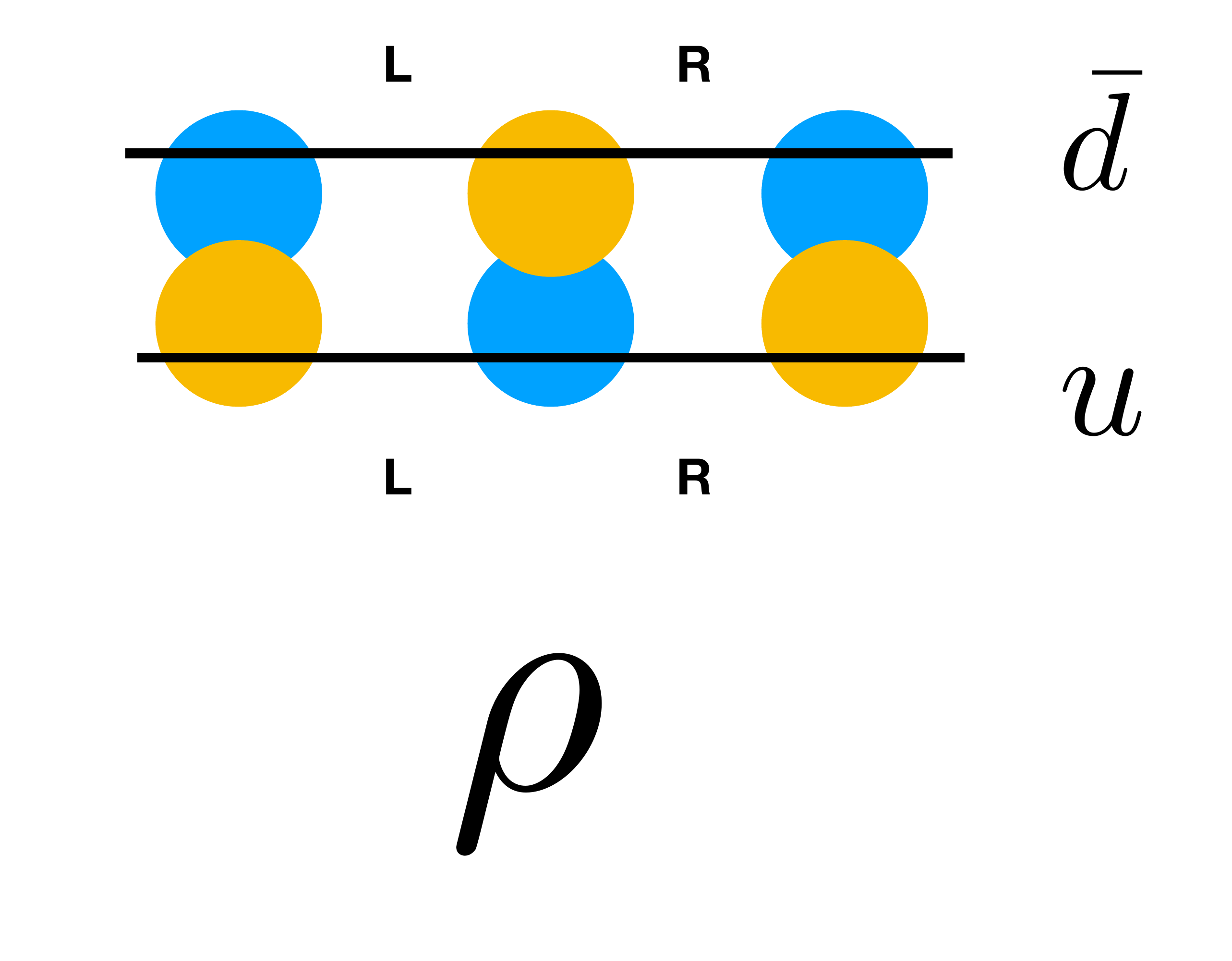}
	\caption{
	Schematic representation of the role of instanton-induced 't Hooft Lagrangian in
	the structure of the pion (upper) and rho meson (lower plot).  }
	\label{fig:pi_rho}
\end{figure}

 However,  for spin splittings in   light quark states these forces coming from Wilson lines
 are not sufficient. The resolution comes from 
 additional terms in the quark propagator,  related to their
 disappearance/creation in/from the Dirac sea  technically  described by parts of the propagators related to {\em fermionic zero modes}. In other words, there are spin forces
induced by 't Hooft effective Lagrangian for light quarks.
In Fig.\ref{fig:pi_rho} we show schematically how these vertices contribute to the pion and rho meson.

\subsection{QCD vacuum and Instanton Liquid Model}

Nonperturbative physics of strong interactions started before the development of QCD in 1970's.
 Nambu and Jona-Lasinio (NJL) \cite{Nambu:1961tp}, inspired by BCS theory of superconductivity, have  qualitatively explained that
 strong enough attraction of quarks can break  $SU(N_f)_A$ chiral symmetry spontaneously and, among many other effects, create
  near-massless pions. Chiral effective Lagrangians and related theory have  lead to one important input,
  a nonzero {\em quark condensate} $\langle \bar q q \rangle\neq 0$.

The discussion of the
``QCD vacuum structure"  started with the QCD sum rules \cite{Shifman:1978bx}, where
 the short-distance description of {\em Euclidean point-to-point correlation functions} 
 via  the Operator Product Expansion (OPE) was related 
 to their long-distance description in terms of the lightest hadrons. 
 The (Euclidean) correlators of QCD operators  were calculated in terms of the leading
  {\em vacuum condensates}, $\langle G^2 \rangle, \langle \bar q q \rangle,$ etc. Yet it was 
  soon realized that the strongest nonperturbative effects (in the scalar and pseudoscalar
 channels) are $not$ described by the ``mean fied" OPE predictions~\cite{Novikov:1981xi}.

Extensive studies of point-to-point (Euclidean) correlation functions in multiple mesonic channels, based on phenomenology 
\cite{Shuryak:1993kg}, have indeed found striking differences between correlators 
of operators with different quantum numbers. While for
vector currents $(\bar q \gamma_\mu q)$ made of $light$   quarks $q=u,d,s$ only small deviations from free quark propagation is observed, correlators of
the scalar and pseudoscalar operators ($\gamma_\mu \rightarrow 1,\gamma_5$) show strong splittings from them at surprisingly small distances. As shown in detail in \cite{Shuryak:1993kg},
these splittings found their explanation in terms of  topological fluctuations
of gluonic fields, described  semiclassically
by instantons \cite{Belavin:1975fg}.

The semiclassical model of the QCD vacuum structure based on instantons \cite{Shuryak:1981fza} is known as the  "instanton liquid model" (ILM)
. It introduced an important scale parameter of the nonperturbative vacuum -- the  typical instanton size{
\begin{equation} \rho\sim \frac 13\, {\rm fm} \label{eqn_size}  \end{equation} 
Because it is rather small compared to  sizes of most hadrons (except for the lowest $\Upsilon$s and the pion, see below), where possible we will use a {\em quasi-local}
approximation in which it is going to zero.

 In the ILM  the vacuum fields are very inhomogeneous:  blobs of 
strong gauge fields (at the instanton centers)  surrounded by ``empty" space-time,
free from nonperturbative fields. Since  mid-1990s such field distributions were 
obtained from lattice configurations by means of various ``cooling" methods.  
 A picture is better than many words: so we reproduce in~Fig.~\ref{fig_VAC} one visualization
 of   the  topological charge distribution  from
 \cite{Leinweber:1999cw,Biddle:2019gke,Biddle:2020eec} which reveals a  number of  instantons and anti-instantons.

 Note that the topological clusters are also threaded by thin center vortices 
or Z$_3$-fluxed strings,  which span  world-sheet surfaces in 4-dimensions. 
 While the center vortices are important for enforcing confinement at long distances, Fig.~\ref{fig_VAC}  shows that they are on average 
decoupled from the inhomogeneous and strong topological fields. Moreover, their field strength in the vicinity of these topological fields is
about $\sigma_T\bar \rho\approx 0.3\,{\rm GeV}$, which is weaker than the typical chromo-electric or chromo-magnetic field
in the instanton center $\sqrt{E}=\sqrt{B}\approx 2.5/\bar \rho\approx 1.5\,{\rm GeV}$,
but crucial for long-distance color correlations. 

\begin{figure}[t!]
	\begin{center}
		\includegraphics[width=9cm]{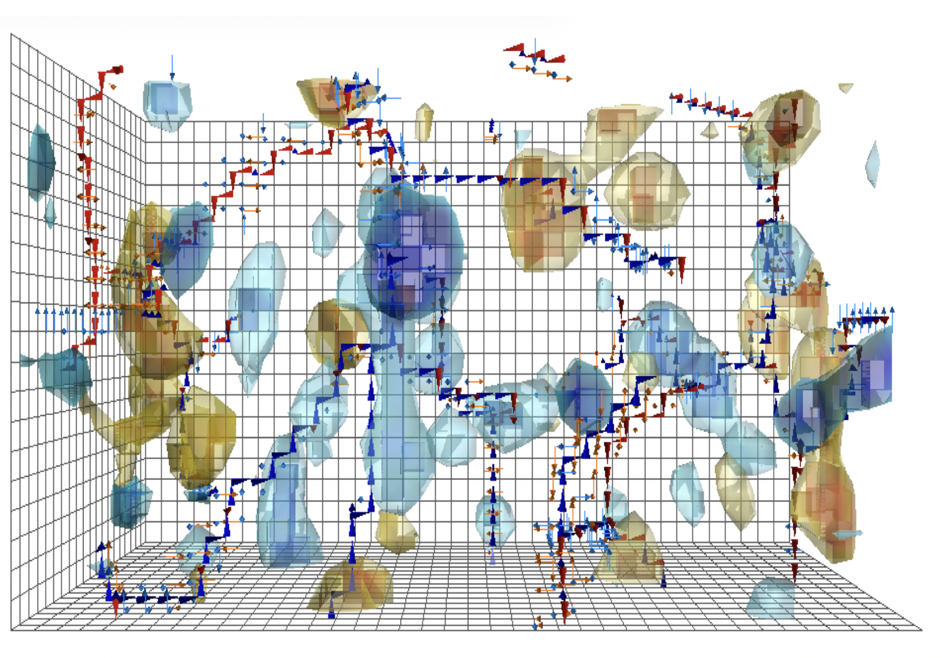}
		\caption{Instantons (yellow) and anti-instantons (blue) configurations in the ``deep-cooled" Yang-Mills  vacuum. After center projection,  they are  threaded by center P-vortices~\cite{Biddle:2019gke,Biddle:2020eec}.
		They constitute the primordial gluon epoxy (hard glue) at the origin of  the light hadron masses~\cite{Zahed:2021fxk}. The center P-vortices are  responsible for confinement. See text.}
		\label{fig_VAC}
	\end{center}
\end{figure}

Instantons are not just semiclassical solitons made of glue. As discovered by   't Hooft  \cite{tHooft:1976snw},
they generate  4-dimensional fermionic zero modes which lead to multifermion
effective Lagrangian. 
The instanton-induced 4-quark
not only solves the famous ``$U_A(1)$ problem" -- by making the $\eta'$ non-Goldstone and heavy --   but it also produces
a strong attraction in the $\sigma$ and $\pi$ channels. Including those
in the framework of the ILM \cite{Shuryak:1982hk} one gets
 a microscopic understanding of
  chiral symmetry breaking,  chiral perturbation theory,  the pion properties etc.
'T Hooft instanton-induced effective Lagrangian is a QCD substitute to the  hypothetical Nambu-Jona-Lasinio \cite{Nambu:1961tp} four-quark interaction. 

Another version of this theory look at
  quarks ``hopping" from one instanton to another. The
resulting loops in the quark determinant can be either short or  long (infinite in thermodynamic limit $V_4\rightarrow \infty$).
  The latter component is responsible for the localization of the Dirac eigenvalues close to zero $\lambda\sim 1/V_4$, 
  
    Two parameters of the original ILM  are the size and the density of instantons
  \begin{equation} \label{eqn_n_rho}
  	 \rho=\frac 13 \,{\rm  fm},\qquad n_{I+\bar{I}}=\frac 1{R^4}=1 \, {\rm fm}^{-4}
 \end{equation}
  which in turn defines the so called {\em diluteness parameter}
   \begin{equation} 
   \kappa\equiv \pi^2 \rho^4 n_{I+\bar I }
 \end{equation}
  So,  the  ILM predicts it to be small $\kappa_{\rm ILM}\sim {\cal O}(1/10)\ll 1$. These parameters 
 have withstood the scrutiny of time, and describe
rather  well the chiral dynamics related to pions, the Euclidean correlation functions in the few femtometers range,
and much more, see~\cite{Schafer:1996wv} for a review.

  Another formulation of chiral symmetry breaking  relates it to  the 
 {\em  collectivization} of the instanton zero modes, into the so-called {\em ``zero mode zone"} (ZMZ).
Due to nonzero matrix elements of the Dirac operator,  near-zero Dirac eigenvalues are residing within a strip of small width
 \begin{equation}  |\lambda | \sim  {\rm width\,(ZMZ)}\sim {\rho^2\over R^3} \sim 20\, {\rm MeV}
 \end{equation} 
as predicted by the original ILM with the parameters given above (\ref{eqn_n_rho}). In~\cite{Glozman:2012fj} the
meson and baryon spectroscopy was studied, with all
Dirac states inside a certain strip of eigenvalues $|\lambda|<\Delta$ eliminated. A strong
restructuring of the light hadronic spectra was indeed observed if $\Delta > {\rm width\,(ZMZ)}$. 
In particular, the Nambu-Goldstone modes (pions)
totally disappear from the spectra, as expected.
It would be interesting to extend the same analysis to the heavy-light  sectors, and to
address more generally what happens with all forms of spin-dependent forces.

Further statistical description of interacting  instanton ensembles was developed
by mean field methods and statistical simulations,
 for review see~\cite{Schafer:1996wv}. Some related 
lattice studies  are \cite{Chu:1994vi,Faccioli:2003qz}.

 \subsection{Topological landscape and $I\bar I$ molecules }
  The ``landscape" refers to  the {\em minimal energy} gauge field configurations, as a function of two main variables. 
The first is the topological Chern-Simons number
\be N_{CS}\equiv { \epsilon^{\alpha\beta\gamma} \over 16\pi^2}\int d^3x  \left( A^a_\alpha \partial_\beta A^a_\gamma +{1\over 3}\epsilon^{abc}A^a_\alpha A^b_\beta A^c_\gamma \right)
	 \label{eqn_Ncs}\ee
 For the second we will use the r.m.s. size of the configuration, related to the field strength
 \be \rho_{r.m.s.}={\int r^2 G^2 d^3 r \over   \int G^2 d^3 r} \ee 
For fixed $\rho$, the landscape is shown in Fig.\ref{fig_landscape}.
 It has been defined in Ref \cite{Ostrovsky:2002cg}  and is given in a parametric form
	 \begin{eqnarray}
	 	U_{\rm min}(k, \rho)&=&(1-k^2)^2{3\pi^2\over g^2\rho} \\ \nonumber
	 	{N}_{CS}(\kappa)&=&\frac 14 {\rm sign}(k)(1-|k|)^2(2+|k|) 	
	 \end{eqnarray}	
with parameter $\kappa$. At $\kappa=\pm 1$ the minimal energy is zero,
while $k=0$ corresponds to a maximum of the energy. This point was called
"the sphaleron"  in electroweak theory 
 \cite{Klinkhamer:1984di}. 
 The set of configurations at arbitrary $\kappa$ is called the {\em sphaleron path}. Those consist of  static 3d magnetic field configurations,
 known also as  the {\em turning points}  (by analogy to points in quantum mechanics where the 
 semiclassical momentum vanishes).
 
 \begin{widetext}
 
 \begin{figure}[h!]
\begin{center}
\includegraphics[width=14.cm]{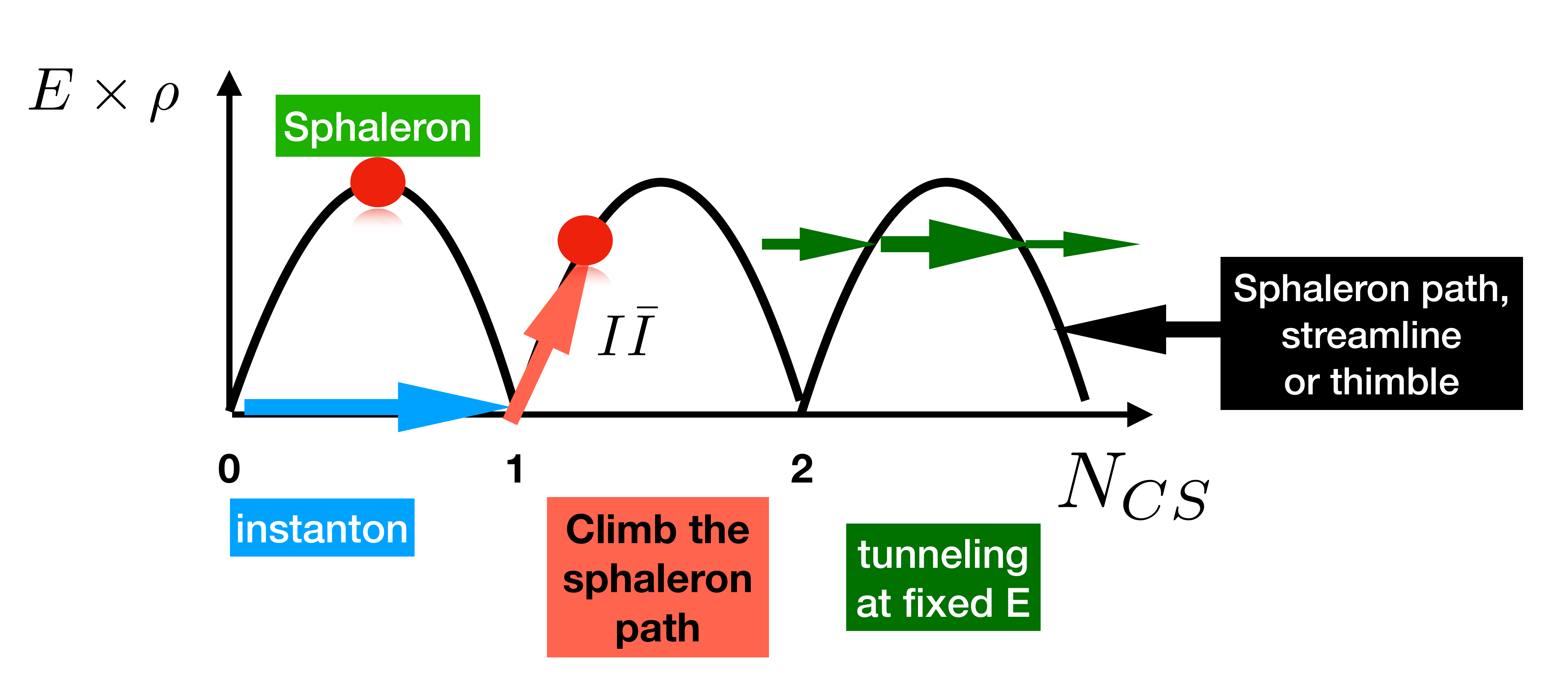}
\caption{The topological landscape:  minimal energy (times r.m.s. size) versus Chern-Simons number $N_{CS}$ (left);
thimble path (center);  tunneling at nonzero energy (right). }
\label{fig_landscape}
\end{center}
\end{figure}

\end{widetext}

 The (anti)instanton is a tunneling path connecting the $bottoms$ of two subsequent valleys, at energy zero.
 Since they have topological charge $Q=\pm 1$, they result in change of Chern-Simons number $\Delta N_{CS}=\pm 1$. 
  
 However instanton paths are {\em not the only form of topological fluctuations} 
 which may occur in this landscape. Indeed, we will discuss $two$ additional  sets of
 paths:
 (i) those that travel the landscape along the sphaleron path  (or $^\prime$streamline$^\prime$);
 (ii)  those that travel the landscape at fixed energy, including the tunneling; 
all of which are  illustrated in Fig.\ref{fig_landscape}
 
 The first set is described by a constrained Yang-Mills equation with a nonzero r.h.s., or external current, 
 which drags them up (or down) the potential along the gradient. In the mathematical literature
this  construction is  known as ``Lefschetz thimbles", a special path connecting two extrema of a function following its gradient. In QCD-related literature they are  known as ``streamline" configurations described by overlapping instanton-antiinstanton pairs or molecules. 
 
 The second set is  described by Yang-Mills equation with a $zero$ r.h.s., and thus 
 is occurring at fixed energy.  The path's history passes the ``turning points " $twice$, with
 Euclidean time solution in between them, known as {\em tunneling at nonzero energy}. In general, those should be complemented by
  Minkowskian time solutions, before and after turning points.   Technical name for those is {\em ``zigzag path"} indicating a turn from real to imaginary time and then back to real.
 
 Both sets of solutions, with proper references,  we detail in the Appendices. What we would like to emphasize here is the fact that they describe 
 gauge field fluctuations {\em different from} well separated instantons. They are neither self-dual  nor possess  (near) zero Dirac eigenvalues -- and therefore they do not  contribute to chiral symmetry breaking and were not included in the original ILM. And still,
   they do describe certain fluctuations of vacuum gauge fields, and therefore can contribute to certain observables:  especially the Wilson lines. Therefore, we will investigate their contributions to
 central and spin-dependent forces between quarks. 
 The inclusion of ``molecules" or "zigzag paths" is the novel element of novel vacuum model we
 develop here.
 
 Close instanton-antiinstanton pairs are of course known, and in particular
 there were  observed on the lattice.  They are however not seen in
 Fig.~\ref{fig_VAC} obtained by so called ``deep cooling"  of configurations,
during which close instanton-antiinstanton pairs are already annihilated.
The application of the ``molecular component"   of the vacuum was made
 previously in connection to phase transitions in hot/dense matter.
Indeed, (if quark masses are neglected) this component is the only one which survives at
 temperatures $T>T_c$, where chiral symmetry is restored. Account for ``atomic" and ``molecular" components together
started with~\cite{Ilgenfritz:1988dh}. The ``molecular component"   was also shown to be
important at  high baryonic densities, where
it contributes to quark pairing and color superconductivity~\cite{Rapp:1997zu}. More recently,
we have explored it in the context through non-perturbative contributions to the mesonic form factors~\cite{Shuryak:2020ktq}
and matching kernels~\cite{Liu:2021evw}.

The theory of sphaleron processes ( sketched  in the middle of Fig.\ref{fig_landscape}. is related to he issue of $I\bar I$ interaction. Numerical calculation of the ``streamline" along the
action gradient was first done for quantum-mechanical instantons in \cite{Shuryak:1987tr}, for gauge theory
the ``streamline equation was derived in \cite{Balitsky:1986qn}, solved approximately by  
\cite{Yung:1987zp} and numerically by \cite{Verbaarschot:1991sq}. The surprising finding
of the latter paper was that ``Yung ansatz" was rather accurate not only at large distances $R\gg \rho$ 
between instanton and antiinstanton, where it was derived, but in fact for any distance till zero. 
Note that the last two papers  used conformal inversion at the center of the instanton,
making $I$ and $\bar I$ co-central. 

Consider $I$ and $\bar I$ of the same size $\rho$ and same color orientation, with 4-distance
between their centers $R$ serving as the parameter of the set, with  $x1_\mu=(0,0,0,R/2),x2_\mu=(0,0,0,-R/2)$. In the mathematical literature such set is known as {\em Lefschets thimble}, it connects one extremum at large $R$, the independent $I$ and $\bar I$,
with another, the zero field at $R=0$. Note that at the locations $y_4<0$ the gauge fields are
approximately antiselfdual $\vec E\approx -\vec B$, and at $y_4>0$ they are approximately
selfdual $\vec E\approx \vec B$. At $y_4=0$ the electric field vanishes. As shown in \cite{Ostrovsky:2002cg}, these 3d magnetic objects obtained using Yung ansatz are very close to the sphaleron path configurations obtained by constrained energy minimization. 

The instanton-antiinstanton streamline therefore provides a semiclassical description of the sphaleron
production. In the context of electroweak theory it was first used in \cite{Khoze:1991sa} and  \cite{Shuryak:1991pn} three decades ago.
Recent interest in the production of QCD sphalerons at
LHC and RHIC colliders is discussed in our recent paper \cite{Shuryak:2021iqu}.

Let us now move to lattice observables, to which
 the ``molecular component" of the instanton ensemble contributes.  The simplest of which is 
the so called {\em gluon condensate}
$\langle G_{\mu\nu}^2 \rangle$ introduced in the context of the QCD sum rules framework~\cite{Shifman:1978bx}.
The accuracy of this number was later questioned and it was revised to a larger value.
Further discrepancies were shown in  lattice studies, that were attempting to extract
 local (or nonlocal)  observables with  powers of 
 the gauge field strength $G_{\mu\nu}$. 
 
 Nowadays, when ``cooling" by the gradient flow method can be  consistently related to the renormalization group (RG) flow \cite{Luscher:2011bx}, one can
 put the appropriate scale dependence of the ``molecular component"  on a firm basis.
Skipping several decades, let us take  as an example the  recent 
work~\cite{Athenodorou:2018jwu}  which studies  the   3- and 4-point gluon field correlators
  and related their evolution to topology  during cooling.
Their original motivation  was to extract the gluon coupling $\alpha_s(k)$,
so the observable was chosen to be 
the  ratio of the 3-point to 2-point Green function leading to effective coupling
\begin{equation}
\alpha_{MOM}(k)={k^6 \over 4\pi} {\langle  G^{(3)}(k^2)  \rangle^2 \over    \langle  G^{(2)}(k^2)    \rangle^3 }  
\label{ratioG3toG2}
\end{equation}
In the ``uncooled" quantum vacuum (with gluons)
 the effective coupling  starts running downward at large  $k>1\, {\rm GeV}$, as rexpected 
by asymptotic freedom. However  at
low $k\rightarrow 0$, one finds another persisting positive power of $k$, with a slope that  matches 
exactly the one following from an instanton ensemble~\cite{Boucaud:2002fx}

\begin{equation}
\alpha_{MOM}(k)\rightarrow \frac{k^4}{18\pi n_{I+\bar I}}
\end{equation}
Furthermore, it was observed that  with increasing cooling
time $\tau$, 
the same power spreads to higher  momenta,   $k> 1\, {\rm GeV}$.  It was observed
that  ``cooling" eliminates not only  perturbative 
gluons, but  close instanton-antiinstanton  pairs as well.
\begin{figure}[htbp]
\begin{center}
\includegraphics[width=7cm]{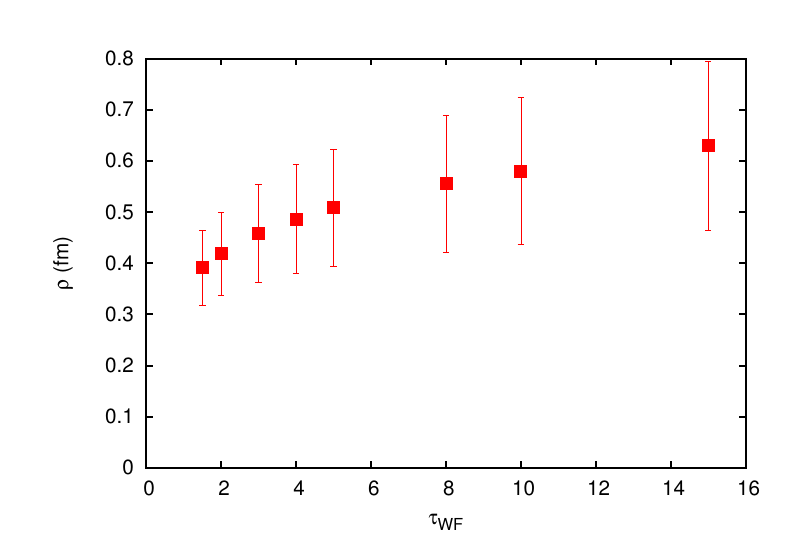}
\includegraphics[width=8cm]{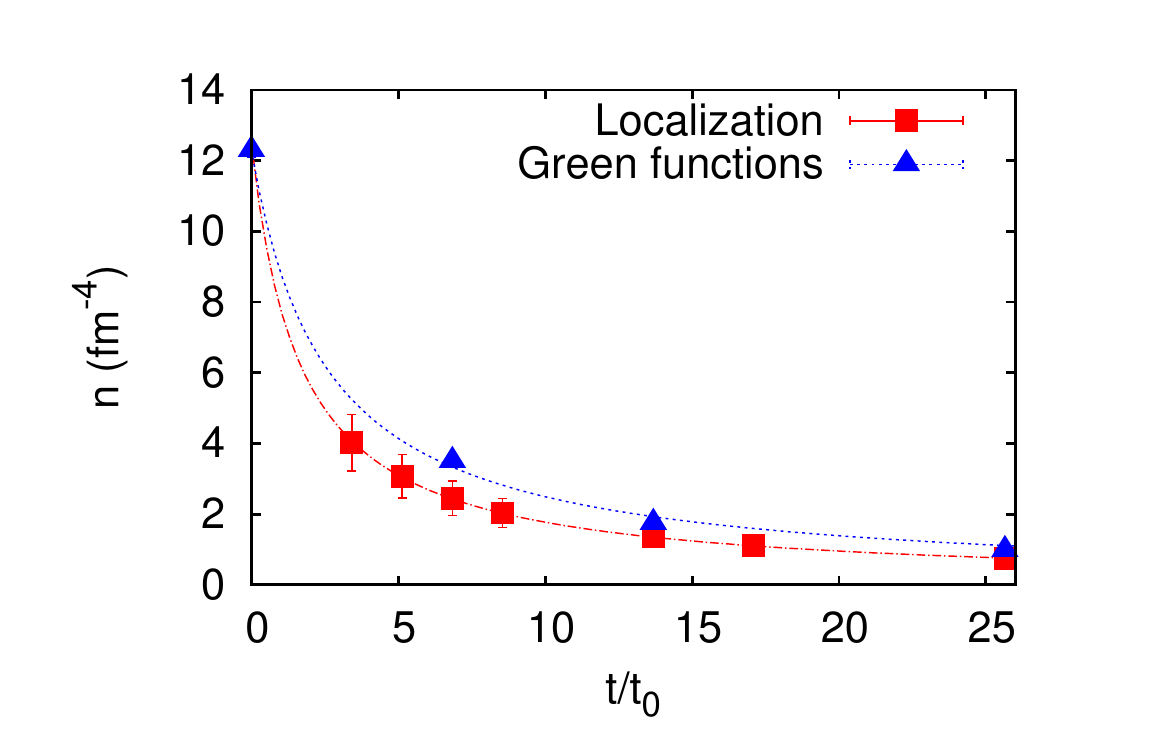}
\caption{The dependence of the mean instanton sizes (upper plot) and density (lower plot)
on the gradient flow cooling time $\tau$ (arbitrary units). The quantum vacuum 
of course corresponds to an extrapolation to $\tau\rightarrow 0$.
}
\label{fig_cooling}
\end{center}
\end{figure}

 The dependence of the mean instanton sizes and density as a function of  the (gradient flow)
 cooling time is shown in Fig.\ref{fig_cooling}, from \cite{Athenodorou:2018jwu} .
 Their main conclusion from this analysis was that the size and density
 -- extrapolated to $zero$ cooling time ($\tau \rightarrow 0$)  --  are
 \be \,\, \rho  \rightarrow {1 \over 3} \, fm, \,\,\,n \rightarrow 10 \, fm^{-4}
 \ee
  Therefor one can see that  
total instanton density is significantly larger than known before from
 ``deep cooling" (large $\tau$) of $1/{\rm fm}^{4}$,  as in the original ILM.
 In fact, the density may even be larger than that, because
 the definition of instantons used in this paper exclude pairs (``molecules") which are 
 too close.
 We will use these numbers, as indicative, for 
 our ``dense instanton liquid model" below.

\subsection{Content of this paper}

 In section \ref{sec_central_dense}
we will argue that the $upper$ limit on the
total instanton density can be deduced from known magnitude of the
central potentials, e.g. from quarkonium spectroscopy. 
%

The main purpose of the present paper is to
quantify {\em nonperturbative  spin-dependent}  forces.  We begin 
by reviewing their phenomenology in section \ref{sec_spin_pheno} and then
the theory in \ref{sec_spindep_theory} which is
based on the standard "Wilson line plus two field strengths" correlators.     The contribution of an instanton to a pair of Wilson lines  can be readily  calculated analytically.
 This approach has been first used to calculate the static $\bar Q Q$ potential by
 the original Princeton group \cite{Callan:1978ye}. Two decades later we generalized it to high energy  $\bar Q Q$ and dipole-dipole scatterings ~\cite{Shuryak:2000df}.
 In the scattering approach,  the  Wilson  lines are first assumed to cross at an angle $\theta_{12}$
 in Euclidean geometry.  The final result is analytically transformed to a  scattering amplitude
(near) the light front by the substitution of the angle $\theta_{12}$  between two Wilson lines,
to the hyperbolic angle $i y$ with  {\em relative rapidity} $y$ between two colliding particles in the ultra-relativistic limit. Unfortunately, neither of these works were
widely noticed   or actually used by phenomenologists (except for  \cite{Giordano:2008ua,Giordano:2009vs}).

An extensive study of the instanton-induced effects on heavy quarkonia has been revived
 recently by Musakhanov et al (see e.g. \cite{Turimov:2016adx}), who calculated
the magnitude of the effect for the central as well as the spin-spin, spin-orbit and tensor forces.  
 Using  the original ILM parameters of \cite{Shuryak:1981ff}, it was found that
 the central potential  is of  magnitude  $\sim 150 \, {\rm MeV}$ at large $r$, a relatively  small correction to  the phenomenological potential.
 The spin-spin potential was also calculated, and was found to be short ranged and of order 
 of about $30\, {\rm MeV}$, also  a  small correction. The spin-orbit and tensor
 forces are even smaller. 
 
 We will modify these results by using our new  model for the instanton ensemble, 
 and compare the results to the  lattice results for spin-spin forces in section \ref{sec_lattice_SS},
  and to the phenomenological vector-pseudoscalar mass splittings in section
 \ref{sec_spindep_theory}.  We will see that in this setting the splittings 
in heavy-heavy quarkonia are reproduced.

The next challenge using the novel model,  is to extend the same analysis  
to heavy-light mesons  such as $B,D...$ and  eventually light-light mesons.
The goal is to find out wether these lighter systems can also be described
analogously using variants of the ``constituent quark" models.

Not all instanton-induced effects are included using temporal
Wilson loops with straight  fermion lines. The part of the quark propagation, related to
the chiral anomaly, instanton zero modes and the  't Hooft Lagrangian has to 
be treated separately as we detail in section~\ref{sec_Hooft}. Only the inclusion of
this  effective interaction can describe pions.

 Yet to get the correct mass and wave function of the vector ($\rho$) mesons, we aldo
 need the effective forces originating from $I\bar I$ molecules, see \ref{sec_rho}.
 The discussion of the heavy-light systems is in section \ref{sec_heavy_light}.
 The conclusions reached  in this study are summarized in section \ref{sec_summary}.

\section{Central  potentials } \label{sec_central_dense}
\subsection{Flux tubes  vs the instantons}

The static quark potential $V_C(r)$ is defined via the  vacuum average of 
the
{\em Wilson line}
  \begin{equation} W=Pexp\big[i g \int  dx^\mu A_\mu^a \hat T^a \big]
  	\end{equation}
over a closed rectangle $r\times T$. In the limit when   the time extent is much larger than the spacial extent  $T\gg r$,
one can ignore 	the small integrals over the spatial direction $r=|\vec x_1 -\vec x_2|$,
and keep only two over the Euclidean time direction, defining the potential by
\begin{equation} e^{-V_C(r) T}=\langle W(\vec x_1) W^+(\vec x_2)\rangle 
\end{equation} 
 For decades this Wilson's definition is used in lattice studies.

%

We already noted that due to quantum vibrations of the QCD string, the flux tube  contribution to $V_C(r)$ is in fact rather uncertain at  the distance scale $\sim 1/3 \, fm$. 
Yet
 precisely this distance range happens to be the most important one for 
 spin-dependent forces.

Another important point is that  the quark spins, due to  the magnetic moments, interact with $magnetic$ fields. The
spin-dependent potentials 
can be defined via the average of two Wilson lines to which  either {\em two magnetic field strengths}
or   {\em product of electric and magnetic fields} are added.
The QCD flux tubes description, however, provides  a description of the $electric$ fields only. Therefore, one might think that they cannot contribute significantly
to the spin forces.

The instantons (antiinstantons) are selfdual (antiselfdual) vacuum fluctuations, in which
the modulus of the magnetic and electric fields are equal. Furthermore,
 the BPTS $instanton$ fields have ``hedgehog" structure $A^a_\mu\sim \eta^a_{\mu\nu} x^\nu$, and since along the straight line  $dx^\mu$  is the same vector, the colors are
 rotated around the same direction. Therefore,  the accumulated rotation angle is given by a
an integral  along the line \cite{Callan:1978ye,Eichten:1980mw}.
  The instanton-induced central potential has the form
  
\begin{equation}
	V_{\rm instanton}(r)={4\pi n_{\bar{I}+I} \rho^3 \over N_c \rho} I\bigg({r\over \rho}\bigg) 
\end{equation}
Here $n_{\bar{I}+I}$ is the instanton plus antiinstanton 4-dimensional density, $\rho$
is the typical instanton size, and
the function $I(x)$ is defined by an integral
over the  location of the instanton center $y_\mu$. 
\begin{widetext}
\bea	
\label{eqn_I_of_x}
 I(x)=\int_0^\infty dy y^2 \int_{-1}^1 dc 
 \bigg[1-  cos(\alpha_1)cos(\alpha_2) 
- {y+x c \over \sqrt{y^2+x^2+2 x y c }} sin(\alpha_1)sin(\alpha_2)\bigg] 
\eea
\end{widetext}
 in which $c$ is the cosine of the angle between $\vec r$ and $\vec y$,  going through the instanton center.
 The two color rotation angles are
 
 \begin{equation}		
 \alpha_1=\pi { y\over \sqrt{y^2+\rho^2}}, \,\,\,	\alpha_2=\pi \sqrt{ y^2+x^2+2 x y c  \over y^2+x^2+2 x y c+\rho^2 } 
 \end{equation}   
 Note that they vanish for zero impact parameter of the line $y=0$, and become $\pi$ if  their impact parameter gets large. In the former case the integrand above vanishes
 as $cos(\alpha)=1$, in the latter $cos(\alpha)=-1$ and the integrand is maximal.
 
%
Using the instanton parameters of the ILM (\ref{eqn_n_rho}) one finds the large distance value of about 150 \, MeV
\cite{Turimov:2016adx}. 
 We however extend  the original ILM
 including also  a ``molecular  $I\bar I$ component".  As a first approximation (to be improved later)
 we will do so by just enhancing the density, keeping the function
 with the same instanton size $\rho=1/3\, {\rm fm}$ (as justified by lattice data of  Fig.\ref{fig_cooling}(a) ).

 The  ``molecular density" $n_{\rm mol}$ should be limited from above 
 by the phenomenological value of the central potential.
The instanton contribution to $V_C(r)$ is shown in  Fig.\ref{fig_two_potentials}  (solid line) for 
 \begin{equation} n\equiv n_{\rm mol}+n_{\rm ILM}= 7. \, {\rm fm}^{-4}   \end{equation}
 Note that this density is about twice lower than indicated by a triangle and extrapolation lines in Fig.\ref{fig_cooling}(b).
 
If so, the instanton-induced central potential gets quite close to the phenomenologically known one,  for intermediate distances $r<  3 \, {\rm GeV}^{-1}$. 
Assuming it is indeed the case, we
 proceed to calculate the spin-dependent effects  below,
and will show that the results are quite reasonable. 

  Note further that with  this choice,  the diluteness parameter  $\kappa\approx 1$. In other words, the ensemble 
is  very dense, with  the mean interparticle distance as low as
$$ R_{\rm dense}\equiv n^{-1/4}=0.61 \, {\rm fm}\approx 2\rho $$
(It may appear  too dense: but remember that the $I\bar I$ molecules are not in fact two independent
instantons, and their fields and action are partially cancelling each other.)

 \begin{figure}[h]
\begin{center}
\includegraphics[width=6cm]{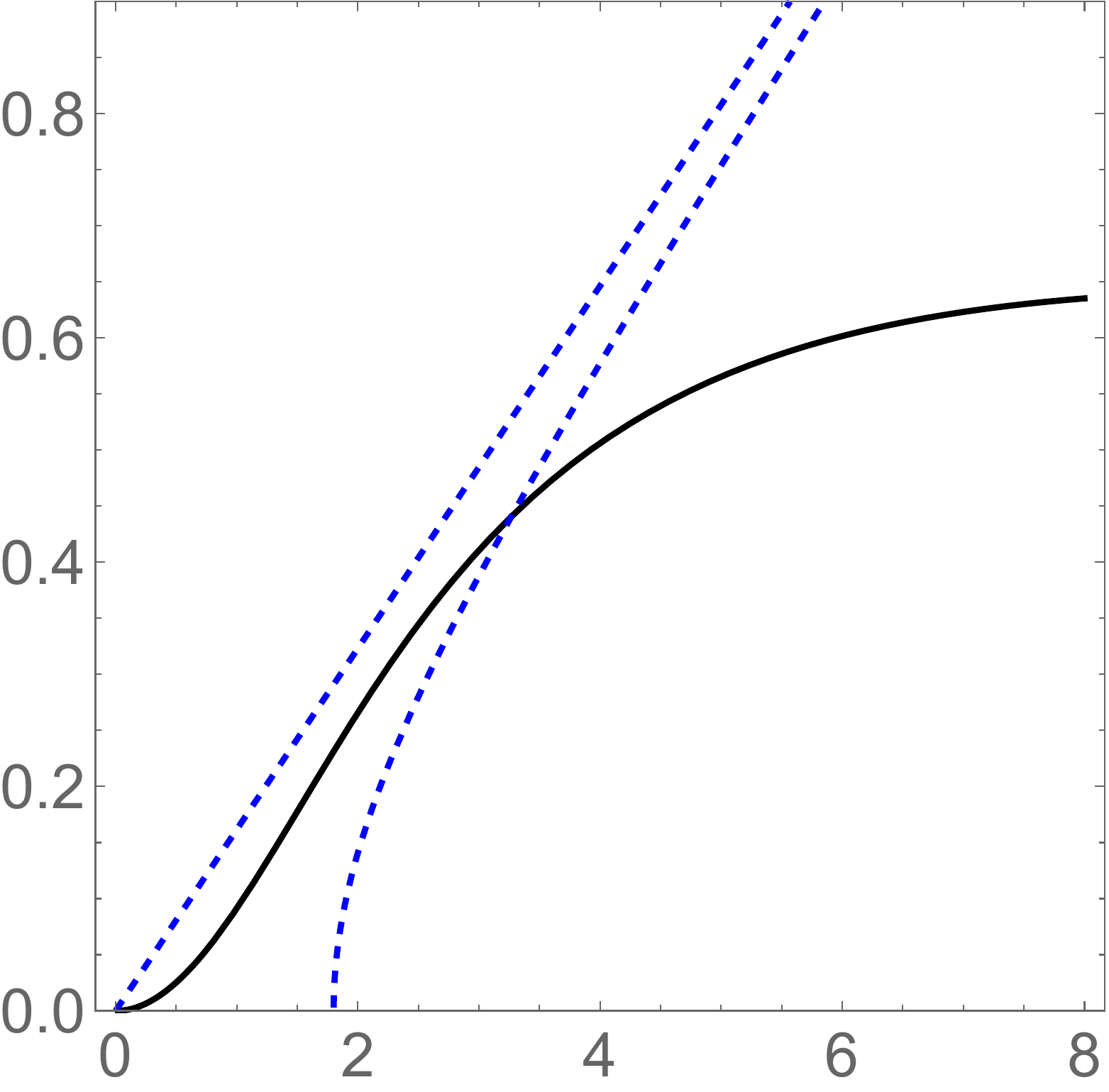}
\caption{Comparison of the central potentials $V_C(r)$
versus the inter-quark distance  $r$ in ${\rm GeV}^{-1}$
(so $1 \, {\rm fm}\approx 5 {\rm GeV}^{-1}$ on this plot).
The solid line is the one derived from the ``dense" instanton ensemble. 
The potential produced by
the confining flux tube  is illustrated by two blue dashed lines, the upper  linear  line is the classical string,  and the lower 
is the Arvis potential  (\ref{eqn_Arvis}) acounting for quantum string vibrations \cite{Arvis:1983fp} . }
\label{fig_two_potentials}
\end{center}
\end{figure}

But before we do so, let us take a well studied example -- the bottonium states -- and 
check to what extent the difference between the linear and instanton-induced potential 
 at $r>1\, {\rm fm}$ is reflected in the mass spectra.
We have calculated the levels of bottomonium $\bar b b$ states using both potentials.
In Fig.\ref{fig_four_upsilons} we show their non-relativistic energies for four radial excitations, with  $S$-shell, orbital momentum
$L=0$. (As the spin-dependent forces were
not included -- we return to those in the next subsections--  therefore the calculation 
corresponds to the {\em spin averaged} combination, of $J=1$ Upsilons and $J=0$ $\eta_b$ states.) 
The difference between these two potentials at large distances 
$r> 5 \, {\rm GeV}^{-1}\approx 1 \,{\rm  fm}$ does indeed translates into  different predictions for  
radially excited states.
At large $n$, the instanton-induced version does not keep
with the expected Regge behavior $m_n^2\sim n$. 
However,  for the select bottomia considered  
 there is little practical difference between the potentials used. 
 (In the matrix elements used for lighter systems
 below,  we will still use the wave functions obtained with the usual Cornell potential.)

 \begin{figure}[h]
\begin{center}
\includegraphics[width=5cm]{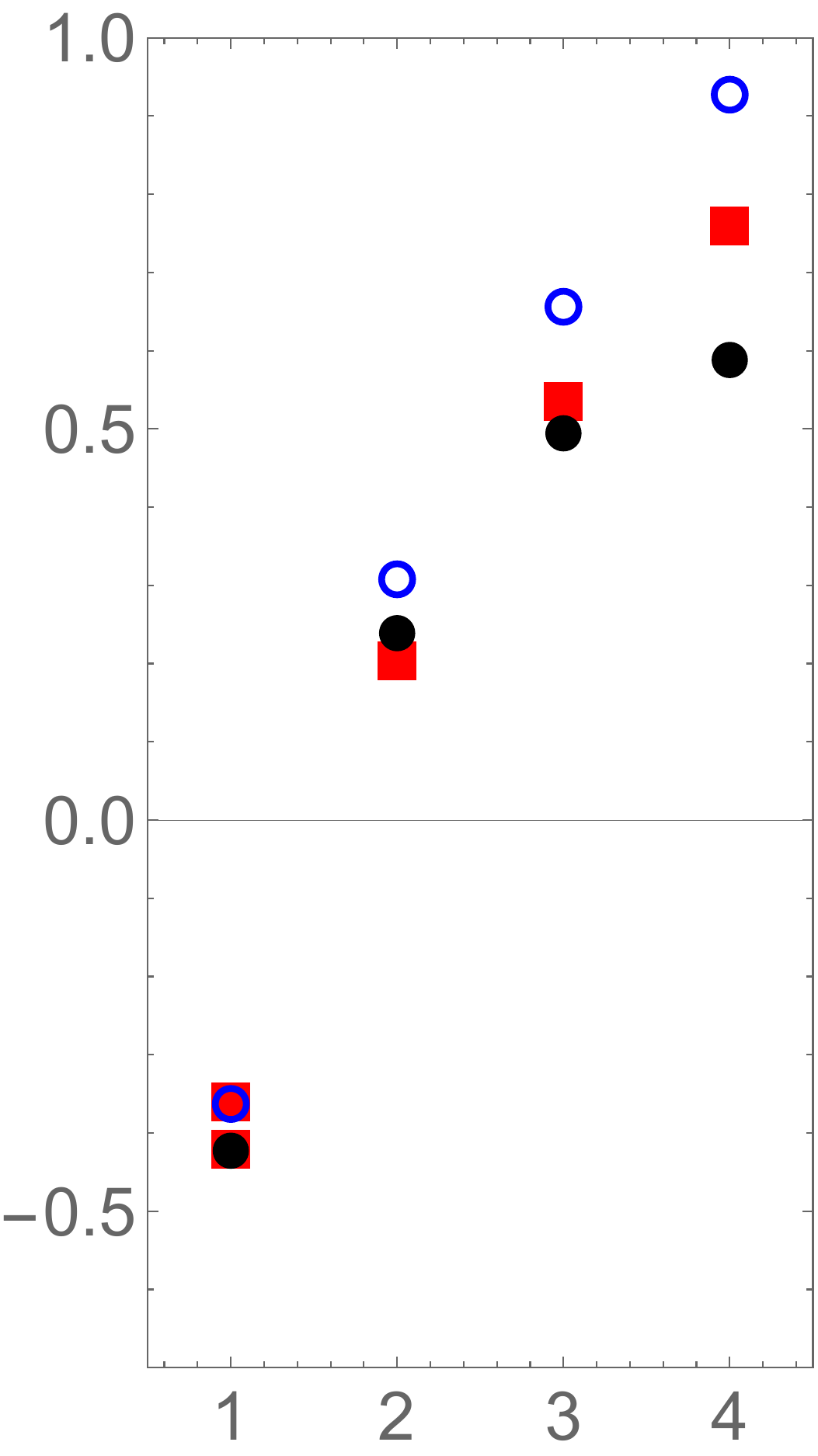}
\caption{Nonrelativisitc energies $M_i-2M_b$ for spin averaged $\bar b b$ states
as a function of the principal quantum number $n$. Five red squares show the experimentally determined masses of $\Upsilon[1S] ,\eta_b[1S], \Upsilon[2S] ,\Upsilon[3S] ,\Upsilon[4S] $, the blue circles correspond to the standard Cornell potential, while the black closed circles are based on the instanton-induced potential shown in Fig.\ref{fig_two_potentials}.}
\label{fig_four_upsilons}
\end{center}
\end{figure}

\subsection{Contributions to $V_C$ of $I\bar I$ molecules} \label{central_molecules}
We start with $I\bar I$ pairs, with a fixed distance $R_\mu$ between the centers. 
The contribution to the central potential strongly depends on the orientation of this vector.
When it is lined  in time direction, $R_\mu=(0,0,0,R)$, as shown in the left of Fig.\ref{fig_molec}, 
the electric field and potential $\vec E(x), A_4(x)$
are time-odd,
  with the opposite sign between lower and upper sub-volumes, the instanton and antiinstanton.
Therefore, the color rotation angles in the Wilson lines -- integrated over time along the lines with arrows --
  get contributions of opposite sign which  $cancel$. We thus come
to the somewhat surprising conclusion that this configuration would  $not$ contribute
to the central potential. 

\begin{figure}[htbp]
\begin{center}
\includegraphics[width=7cm]{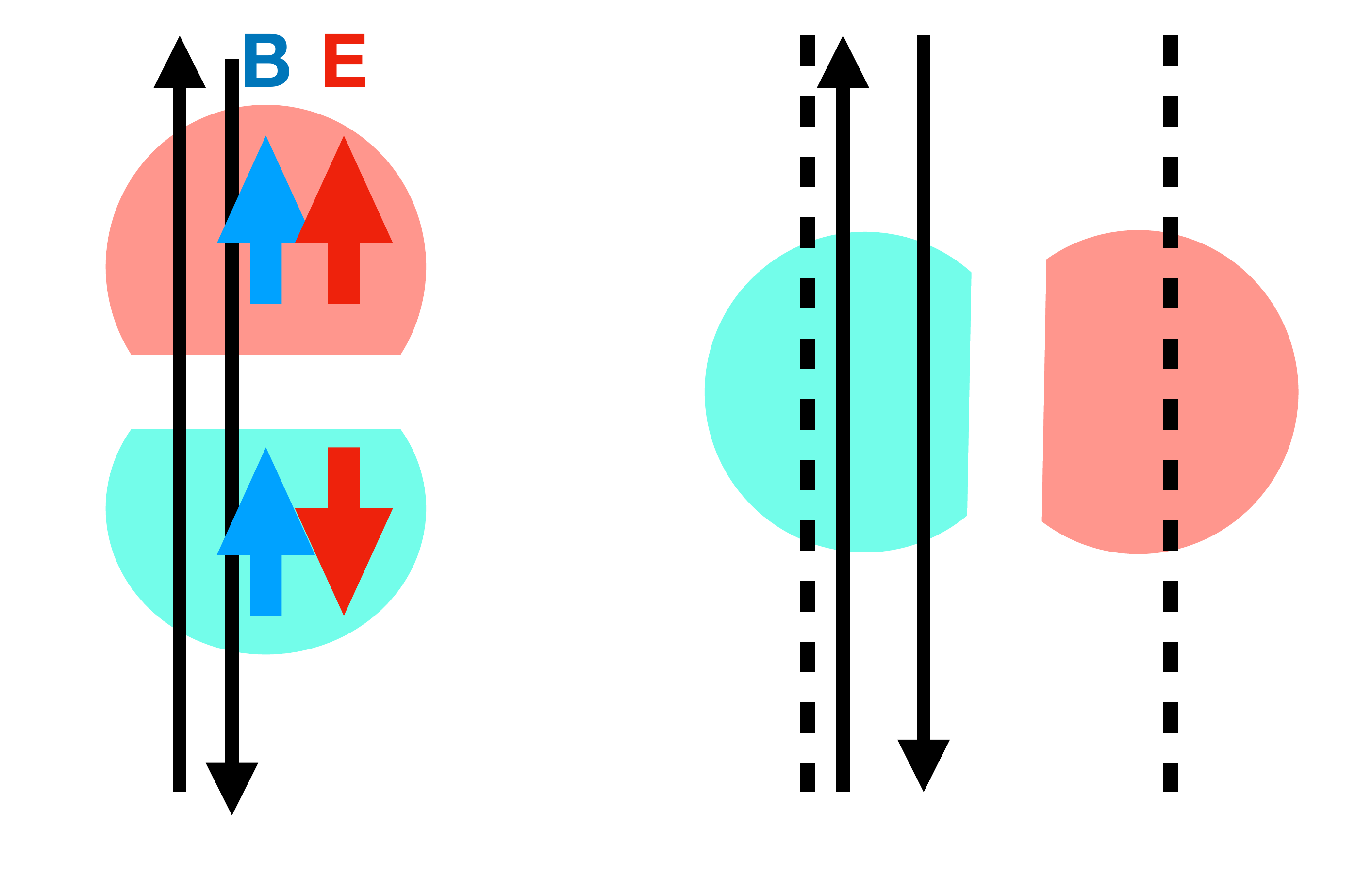}
\caption{Instanton-antiinstanton molecule oriented along  the time axes (left) and space axes (right). }
\label{fig_molec}
\end{center}
\end{figure}

This is not so if $R_\mu$ has other orientations, e.g. in one of the space direction. 
Then the  time-oriented Wilson lines would go through either the self-dual or anti-selfdual parts of the ``molecule".
At small enough distance between both of them $r<\rho$  they are likely to go through the same
duality, and therefore the result would be qualitatively the same as for a single instanton (discussed
in the preceding subsection). At large $r\sim 2\rho$ the opposite would be dominant, with 
one of the angles in expression (\ref{eqn_I_of_x}) changing sign.

\subsection{Fixed energy tunneling events and their contribution to $V_C$}
Tunneling in vacuum in topological landscape proceed at energy zero.
But for quarks moving inside hadrons, or during some hadronic reactions, a certain amount of kinetic energy is available.
 Therefore tunneling through topological barriers may occur not via BPST instanton  but 
 with modified tunneling solution,
at some nonzero energy.
These paths were sketched in the right side of Fig.\ref{fig_landscape}: 
they include two Minkowskian sections of the path, and an Euclidean one. The analytic
solutions for a family of such paths, of  
Yang-Mills equations, can be obtained. 

Their
Euclidean part  interpolates between
two ``turning point" magnetic configurations. As a result, the change in Chern-Simons number is fractional. Depending on the energy, one can obtain a single-parameter family of solutions, which interpolate  between
the bottom of the topological barrier (instanton) and its top (the sphaleron configuration). When analytically continued to Minkowski
signature they  describe implosion and explosion to and from the magnetic turning points.  

{\em Step 1} to obtain such solutions  use the O(4)  static and symmetric
ansatz for the SU(2) configuration in regular gauge

\be
A^a_{ M} (y) = 2\,\bar\eta_{a MN}\,\frac{y_N}{y^2}f(\xi)
\label{Z1}
\ee
with the conformal variable $\xi(y)=\frac 12 {\rm ln} (y^2/\rho^2)$. The passage to 
 singular gauge follows from the substitution $f(\xi)\rightarrow f(\xi)-1$. The corresponding 
 gauge invariant action is
\be
\label{Z2X}
S_S=-\frac 14\int d^4y\,{F}^a_{MN}{ F}^a_{MN} 
\ee
or in terms of the conformal variable

\bea
S_S={24\pi^2 }\int d\xi\left(\frac {f^{\prime 2}(\xi)}2 +V(f(\xi))\right)
\eea
with the inverted double well potential 
$$-V(f)=-2f ^2(1-f)^2$$
in Euclidean signature.
The O(4) profile $f(\xi)$ extremizes (\ref{Z2X}) and solves the Jacobi equation

\be
\frac{d^2f}{d\xi^2} = 4(f^2-f) (2f-1)\,\,.
\label{Z3}
\ee
The solution to (\ref{Z3}) with a sphaleron-like turning point  at $\xi=0$
with zero {\it momentum} $f^\prime(\xi=0)=0$, is

\begin{widetext}
\be
\label{Z4X}
f_k(\xi)=\frac 12\bigg(1+\left(\frac{2k^2}{1+k^2}\right)^{\frac 12}{\rm sn}
\bigg(\xi \bigg(\frac 2{1+k^2}\bigg)^{\frac 12}-K(k), k\bigg)
\ee
\end{widetext}
with $sn$ the Jacobi sine function.
(\ref{Z4X}) forms a k-family of 
$\xi$-{\rm periodic} functions with the period

\be
\label{Z7}
T_k=4K(k)\left(\frac{1+k^2}2\right)^{\frac 12}
\ee
with  $K(k)$ the elliptic function.  

The solutions (\ref{Z4X}) carries  {\it energy} (conjugate to $\xi$-coordinate) at the turning point

\be
\label{Z5}
E_k=\frac{24\pi^2}{\rho}V(f_k(\xi=0))=E_0\bigg(\frac{1-k^2}{1+k^2}\bigg)^2 
\ee
with $E_0=3\pi^2/\rho$.
(\ref{Z5})  interpolates continuously between the sphaleron and the instanton for $0\leq k\leq 1$.
At  $k=0$  the  energy $E_0$ is maximal and corresponds to sphaleron mass,
 the period is $T_0=\sqrt 2 \pi$ and
the profile $f_0(\xi)=\frac 12$ is constant. At $k=1$ the instanton energy is $E_1=0$, the period 
$T_1=\infty$ and the profile is

\be
\label{Z6}
f_1(\xi)=\frac{1}{2}+\frac{1}{2}{\rm sn}(\xi-K(1),1)=\frac{e^{2(\xi-K(1))}}{1+e^{2(\xi-K(1))}}
\ee
Note that the argument shift with $K(1)$ in (\ref{Z6}),  amounts to a rescaling of the instanton size in the conformal coordinate
$\rho\rightarrow \rho\,e^{K(1)}$.

In general, the solution
(\ref{Z4X}) carries Chern-Simons number $N_{k}$ as well, that is tied to the
energy $E_k$ through the profile of the potential 

\be
\label{Z6X}
\left(\frac{E_k}{E_0}\right)=16N^2_{k}(1-N_{k})^2
\ee
with $N_{k=1}=1$  the instanton topological charge, and $N_{k=0}=\frac 12$  the sphaleron Chern-Simons number.
Only the solution with $N_1=1$ is self-dual and stable. All the other
sphaleron-like configurations with $N_k<1$ are unstable  extrema. The flavor analogue of these  configurations were recently used to 
construct stable holographic tetraquark states~\cite{Liu:2019mxw,Liu:2019yye}.

{\em Step 2} is applying the conformal (stereographic) mapping of the previous solutions
that projects
the O(4) turning points on the sphere at $\xi^2=\rho^2$ onto the planar turning point  at $t_E=0$,

\bea
\label{MAP1}
(x+a)_\mu={2 \rho^2 \over (y+a)^2} (y+a)_\mu
\eea
with $a_\mu=(0,0, 0, \rho) $. As a result, the gauge fields depend separately on the Euclidean coordinates $t_E=x_4-Z_4$ and
$r=|\vec x -\vec Z|$ (and  parametrically on $k$).

\begin{widetext}
\bea
\label{MAP2}
&&A^a_4(t_E, r; k)= \bigg({ 8\rho\,t_E x_a \over [(t_E+\rho)^2+r^2]  [(t_E-\rho)^2+r^2]  } \bigg) f_k(\xi_E)\nonumber\\
&&A^a_i(t_E, r; k) =\bigg({ \delta_{ai}(-t_E^2-r^2+\rho^2)+2\rho \epsilon_{aij} x_j +2 x_i x_a \over [(t_E+\rho)^2+r^2]  [(t_E-\rho)^2+r^2]  }\bigg)4\rho f_k(\xi_E)\nonumber\\
\eea
\end{widetext}
with $Z_\mu=(\vec Z, Z_4)$ the collective position, and

\bea
\label{XIE}
\xi_E=\frac 12{\rm ln}\bigg(\frac{(t_E+\rho)^2+r^2}{(t_E-\rho)^2+r^2}\bigg)
\eea
In Euclidean signature,
 $A_4$ and the electric field are real and vanishing at $t_E=0$ as they should,
while  $A_i$ and the magnetic field are finite and real.  
Similar configurations were originally discussed in~\cite{Luscher:1977cw,Schechter:1977qg}.
 The half periodicity (\ref{Z7}) 
maps onto the tunneling time ($r=0$)

\begin{figure}[htbp]
\begin{center}
\includegraphics[width=6cm]{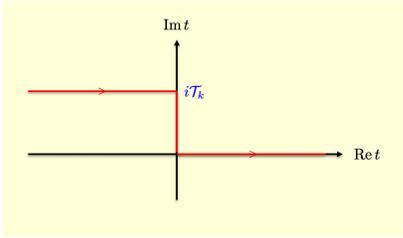}
\caption{Zig-zag path for a temporal  Wilson line describing a heavy-quark traveling from one vacuum to another with a different topological charge, and
riding a fixed energy tunneling configuration with tunneling time $ {\cal T}_k$.}
\label{fig_zig}
\end{center}
\end{figure}

\be
\label{CALTK}
{\cal T}_k=\rho\,\bigg(\frac{e^{\frac 12 T_k}-1}{e^{\frac 12 T_k}+1}\bigg)=
\rho\,{\rm tanh}\bigg(\frac 14 T_k\bigg)\leq \rho
\ee
The conformal transforms of the electric and magnetic fields in (\ref{Z10})  are lengthy. 
The sphaleron configuration with $k=0$ and $f_0=\frac 12$, has $A_4=0$ at the turning point $t_E=0$, with an O(3) symmetric and well localized 
squared magnetic field

\bea
\label{MAP3}
 \vec{B}^2(0,r;0)={96 \rho^4 \over (\rho^2+r^2)^4 }
 \eea

When analytically continued to Minkowski signature $t_E\rightarrow -it$, the gauge fields (\ref{MAP2}) describe a spherically outgoing (incoming)
luminal thin shell, as the exiting sphaleron explodes (implodes) on its way downhill (uphill). The luminal shell supports only light-like fields which
are purely transverse, and fall off as $1/t$ at large times. It was previously used to describe ``sphaleron explosion" in \cite{Ostrovsky:2002cg}.

The fixed energy topological solutions contribute to Wilson lines, as a heavy quark
travels from one vacuum to another vacuum with one added (subtracted) topological charge. In
Minkowski signature, it corresponds to the zig-zag path shown in Fig.~\ref{fig_zig},

\bea
\label{CPATH}
{\cal C}=&&]-\infty+i{\cal T}_k,i{\cal T}_k]\cup [i{\cal T}_k, 0]\cup [0,+\infty[\nonumber\\
\eea
where on the purely imaginary $[ i {\cal T}_k,0]$ path, the heavy quark is riding the tunneling 
process. A similar path choice was also advocated for scattering processes through instantons
in~\cite{Rubakov:1992ec}.

The  chief contribution to the time-like Wilson line follows from this path, since the gauge field
on the paths with real time support is  luminal and transverse asymptotically,

\begin{widetext}
\bea
W(r;k)={\rm exp}\bigg(i\bigg[\int_{-\infty+i {\cal T}_k}^{i {\cal T}_k}dt\,\frac 12A_4^a(-it, r; k)\tau^a+\int_{i{\cal T}_k}^{0}dt\, \frac 12A_4^a(-it, r; k)\tau^a+\int_0^\infty dt\, \frac 12A_4^a(-it, r; k)\tau^a\bigg]\bigg)\nonumber\\
\label{WZIG}
\eea
\end{widetext}
At the sphaleron point with $f_{k=0}=1/2$,  we  can check that  $W(r; k=0)=1$ and no self-energy is generated in going
through a finite energy tunneling configuration. This result follows by  deforming  the contour on the real axis, without 
encountering poles for ${\cal T}_{k=0}<\rho$, and noting that $A_4$ is time-odd.  Away from the sphaleron point, a fraction of the
instanton self-energy can be picked, through the emergence  of complex singularities solution to  $f_k(\xi)=\infty$.

\section{Spin-dependent potentials}
\subsection{Phenomenology,  from heavy to light quarks } \label{sec_spin_pheno}

To start, let us consider the first $1S$-shell,  with zero orbital momentum $L=0$, 
to recall  the magnitude of the  spin-spin forces. Starting from heavy quarkonia and proceeding to
lighter systems,  we focus on the following selection of mass splittings (all  in MeV)
\begin{eqnarray} \label{eqn_SS_splittings}
 M(\Upsilon)-M(\eta_b)&=&61.; \,\,\ \end{eqnarray}
\begin{eqnarray}  M(J/\psi)-M(\eta_c)&=&116.;  \,\,\,  M(\psi^{2S})-M(\eta_c^{2S})= 51.            \nonumber \\
 M(B^*)-M(B)&=& 35.5; \,\,\, \,\,\, M(D^*)-M(D)=137.  \,\,\, \nonumber \\
  M(K^*)-M(K)&=&398.;\,\,\,M(\rho)-M(\pi)=636.  \,\,\, \nonumber
\end{eqnarray}
Note that these splittings grow
for lighter quarks, eventually  getting  comparable to the scale of the mesonic masses: 
therefore in those case spin forces cannot be treated as a perturbation. 

Naively, the flavor dependence of the spin-spin forces should just follow from the product of the quark magnetic moments, which is $
\sim 1/m_{Q_1}m_{Q_2}$, times some universal magnetic fields in the QCD vacuum as
defined in the correlators of section \ref{sec_correlators}.
Yet the select splittings  above show  that it is not the case  even for heavy quarks: e.g. charm and bottom 
effects are different by a factor of two, not 10 as the mass ratio would suggest.
Yet universal
vacuum correlators /potentials do not imply universal (mass-independent) matrix elements, since all these
mesons have vastly different wave functions. Especially interesting is the case of charmonium (second line 
in (\ref{eqn_SS_splittings}) ) in which one has both the splittings of 1S and 2S levels. Not only
the overall volume  and $\psi(r=0)$ are different, but also the wave function of 2S state has a node.
The ratio of wave functions squared is shown in Fig.\ref{fig_2Sto1S}, and one can see that it changes from 
$0.538$ at $r=0$ to zero at $r\approx 2\, GeV^{-1}$.  The ratio of the observed splittings is $51./113. \approx 0.45$ is comfortably in between, but much closer to the former number than to the latter one. This observation alone tells us
that $V_{SS}$ must be concentrated at  very small distances. 

Going further in phenomenology, we now proceed to the 1-P shell, with $L=1$. For any flavor combination, 
there are 4 states  we will be interested in as above, starting 
from heavy quarkonia, to heavy-light and all the way to mesons made of light  quarks. 
For the light mesons  we select two channels, the strange-light $K$ mesons and charge $I=1$ sector
of the light-light. (The  $I=0$ sector has complicated mixing between quark states and glueballs,
which we prefer not to include.)
 The results are listed in Table~\ref{tab_abcX}. For example, the $\chi_{b2}$ state  is the $\bar{b} b $ state with $S=1,L=1,J=2$,
and the $h_b$ state  has $S=0,L=1,J=1$. All the names and masses are from the 2020 PDG tables

\begin{figure}[t]
\begin{center}
\includegraphics[width=6cm]{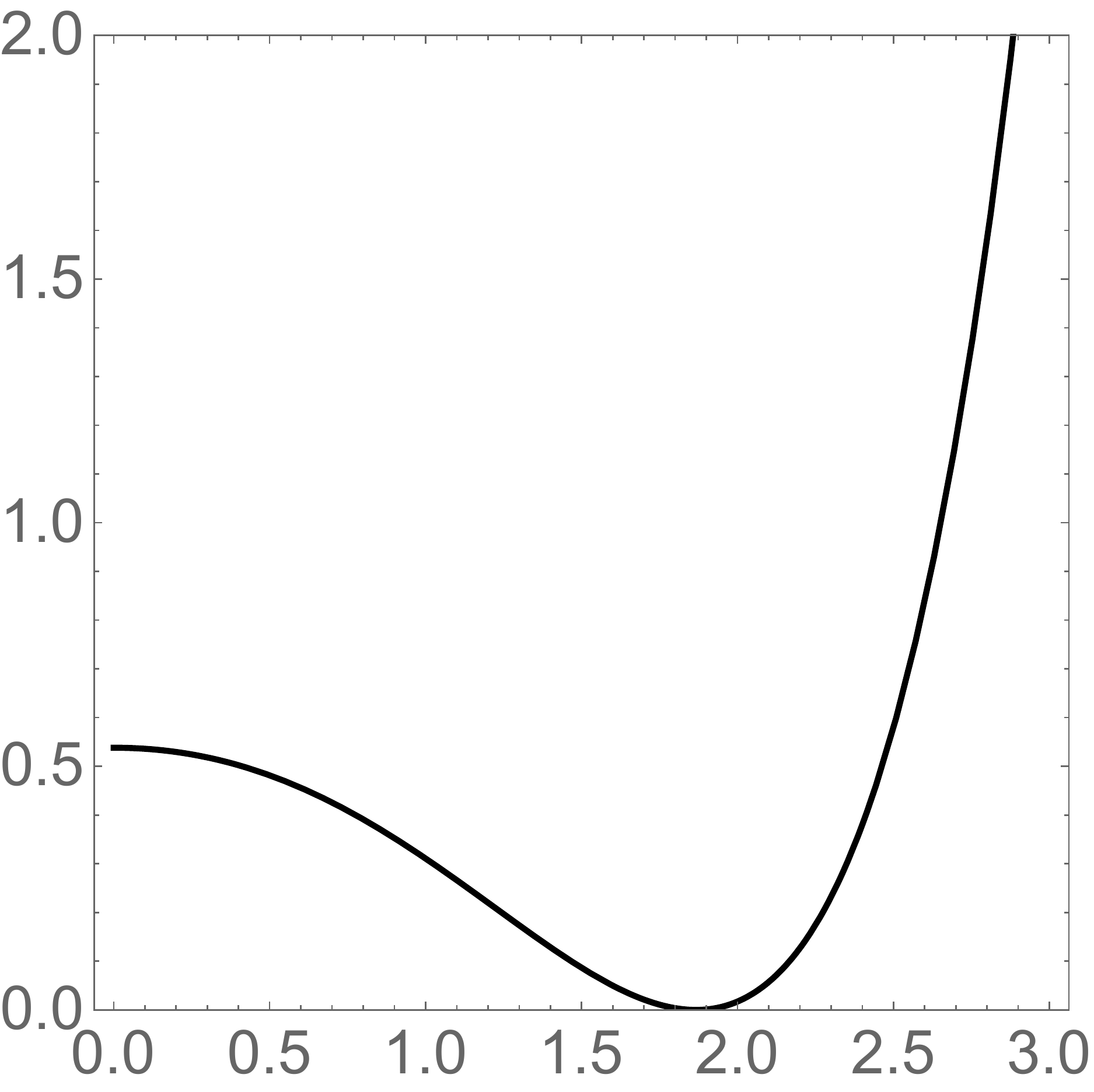}
\caption{The ratio of wave functions squared $\psi_{2S}(r)^2/\psi_{1S}(r)^2$ as a function of distance $r (GeV^{-1})$.}
\label{fig_2Sto1S}
\end{center}
\end{figure}

 With $L=1$ all the three spin-dependent terms come into play, and with four masses one has three differences
which allow  us  to solve for each  individual contribution.  We define the differences from the $\chi$ to $h$ states as 
$$ a \langle (\vec S\cdot \vec L)\rangle  + b \langle S_{12} \rangle +c  \langle (\vec S_1\cdot \vec S_2)\rangle $$
We can solve for  $a,b,c$  by using the corresponding quantum numbers for these states, namely

\begin{eqnarray}
M(\chi_{b2}) - M(h_b) &=&  a - (2/5)b +c,\nonumber \\
M(\chi_{b1}) -M( h_b)  &=& - a + 2b+c,\nonumber \\
M(\chi_{b0}) - M(h_b)  &=&  - 2 a - 4 b+c
\end{eqnarray}
The results  are listed in the last three columns of Table \ref{tab_abcX}.  
The ensuing comments are:
\\
\\
\noindent({\bf i}) The {\em spin-orbit} matrix element $a$ for heavy quarks,  is dominant; \\
({\bf ii}) The  {\em spin-orbit} gets smaller for light quarks.  Why? \\
({\bf iii}) The {\em tensor force} grows for heavy-light but then flips sign for light quarks. Why?\\
({\bf iv}) The {\em spin-spin} term  grows and is  dominant for light quarks.

We now proceed to see whether  these observations can be explained
by the theory we discuss below.

\begin{table}[b!] 
	\caption{The first column is the  flavor composition. The next four columns refer to four states of the $n=1,P$-shell with the quantum numbers explained in the text.
	 The last three columns are
	matrix elements of spin-orbit, tensor and spin-spin terms, in MeV.}
	\begin{center}
		\begin{tabular}{|c|c|c|c|c|c|c|c|}
			\hline
 & S=1,	J=2 &  S=1,J=1 &  S=1,J=0 & S=0,J=1 & a & b & c \\ 	\hline		
 $\bar{b} b$ &	$\chi_{b2} (9912)$ &  $\chi_{b1} (9893)$ & $\chi_{b0} (9859)$ & $h_b(9899)$ & 13.7	& 3.3 & 0.5 \\
  $\bar{c} c$ &	$\chi_{c2} (3556)$ &  $\chi_{c1} (3511)$ & $\chi_{c0} (3415)$ & $h_c(3525)$ & 35.	& 10. & 0.1 \\
     $\bar{c} q$ &	$D_2^* (2461)$ &  $D_1 (2430)$ & $D^*_0 (2300)$ & $D_1(2421)$ & 34.	&  16. & 12. \\
    $\bar{s} q$ &	$K_2 (1430)$ &  $K_1 (1403)$ & $K_0 (1430)$ & $h_s(1270)$ & 6.7	& -5.6 & 151. \\
        $\bar{q} q$ &	$a_2 (1317)$ &  $a_1 (1230)$ & $a_0 (1450)$ & $b_1(1235)$ & 0.5	& -36. & 68. \\
			\hline
		\end{tabular}
	\end{center}
	\label{tab_abcX}
\end{table}%

\subsection{Wilson lines and the five potentials}

Eichten 
and Feinberg~\cite{Eichten:1980mw}   defined general spin-dependent  interactions of heavy quarks  in terms of a priori five  potentials
(see their definitions and further discussion in appendix A) which contribute as follows

\begin{widetext}
\bea
\label{SD}
V_{SD}=&&\bigg(\frac{S_{ Q}\cdot L_Q}{2m^2_Q}-\frac{S_{\bar Q}\cdot L_{\bar Q}}{2m_{\bar Q}^2}\bigg)\bigg(\frac 1r\frac d{dr}\big(V_C(r)+2V_1(r)\big)\bigg)\nonumber\\
+&&\bigg(\frac{S_{\bar Q}\cdot L_Q}{m_Q m_{\bar Q}}-\frac{S_{Q}\cdot L_{\bar Q}}{m_{\bar Q} m_{\bar Q}}\bigg)\bigg(\frac 1r\frac d{dr}V_2(r)\bigg)
+\frac {(3S_{Q}\cdot \hat r S_{\bar Q}\cdot \hat r-S_Q\cdot S_{\bar Q})}{3m_Qm_{\bar Q}}V_3(r)
+\frac 13\frac{S_Q\cdot S_{\bar Q}}{m_Qm_{\bar Q}}V_4(r) 
\eea
\end{widetext}
$\vec S_{Q, \bar Q}$ and $\vec L_{Q, \bar Q}$ are the spin and orbital angular momenta of the $\bar Q Q$ pair. $V(r)$ is the central static potential, $V_1(r)$ and
$V_2(r)$ are obtained by inserting a chromo-electric or chromo-magnetic field on the temporal Wilson loop. The spin-spin and tensor contributions $V_{3,4}(r)$ follow from the insertion of two chromo-magnetic fields on the Wilson loop. (\ref{SD}) is exact to order $1/m_Q^2$.
 The last spin-spin part is fixed as $V_4(r)=2\nabla^2 V_2(r)$~\cite{Eichten:1980mw}.  Lorentz invariance ties the central potential
$V_C(r)$ to  the spin-orbit potentials $V_{1,2}(r)$ through the so called Gromes relation
\cite{Gromes:1984ma}
\be
\label{V12}
V_C(r)=V_2(r)-V_1(r)
\ee
The spin-dependent contributions emerging from the string were
discussed by Buchmuller~\cite{Buchmuller:1981fr} and others~\cite{Pisarski:1987jb,Gromes:1984ma}. Since the spin-spin interactions are short ranged, 
only the spin-orbit contributions survive at large separation $b_\perp$. This is manifest from (\ref{V12}) with $V_2(r)\rightarrow 0$ and $V_C(r)\rightarrow -V_1(r)$ asymptotically,
hence~\cite{Buchmuller:1981fr,Pisarski:1987jb,Gromes:1984ma}

\bea
V_{SL,{\rm string}}(b_\perp)&=&-\bigg(\frac{S_Q\cdot L_Q}{2m_Q^2}-\frac{S_{\bar Q}\cdot L_{\bar Q}}{2m_{\bar Q}^2}\bigg)\frac {\sigma _T}{b_\perp} \nonumber\\
&\rightarrow& -\frac {\sigma_T}{2m_Q^2b_\perp}\,S\cdot L
\eea
with $V_C(r)=\sigma_T r$, $\vec S=\vec S_Q+\vec S_{\bar Q}$ and $\vec L=\vec L_{Q}=-\vec L_{\bar Q}$.
Since the electric flux tube is confined to the string, the spin-orbit contribution is
only due to Thomas precession which is of opposite sign and half the spin-orbit contribution from the cental potential. 
This can be understood from (\ref{SD}-\ref{V12}) if we assume that $V_2(r)$ is short range, so that $V_C(r)=-V_1(r)$.
This string-induced spin-orbit effect 
is dubbed scalar-like in contrast to the Coulomb-induced spin-orbit effect which is vector-like.

\begin{figure}[h]
\begin{center}
\includegraphics[width=6cm]{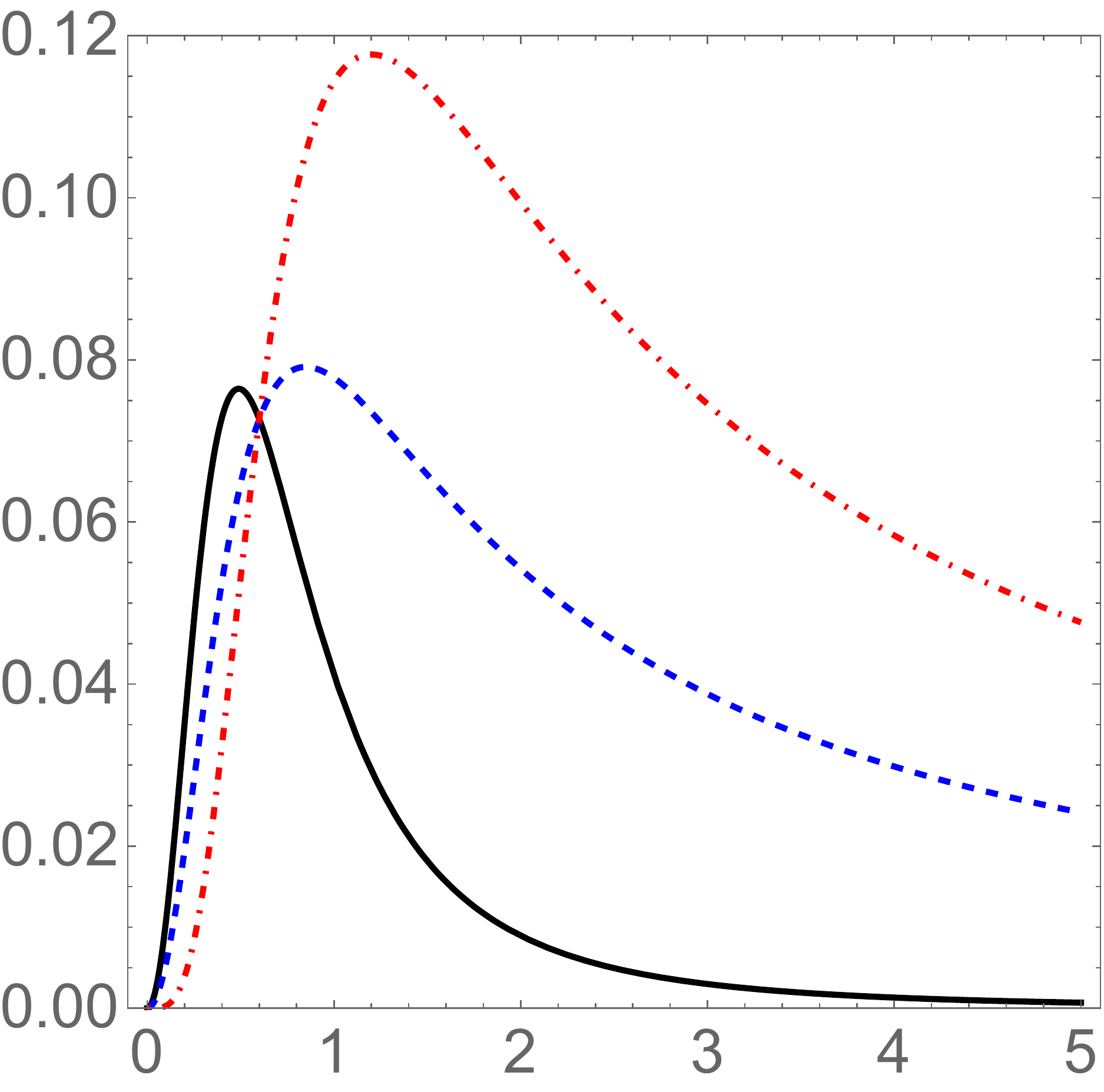} \\
\includegraphics[width=6cm]{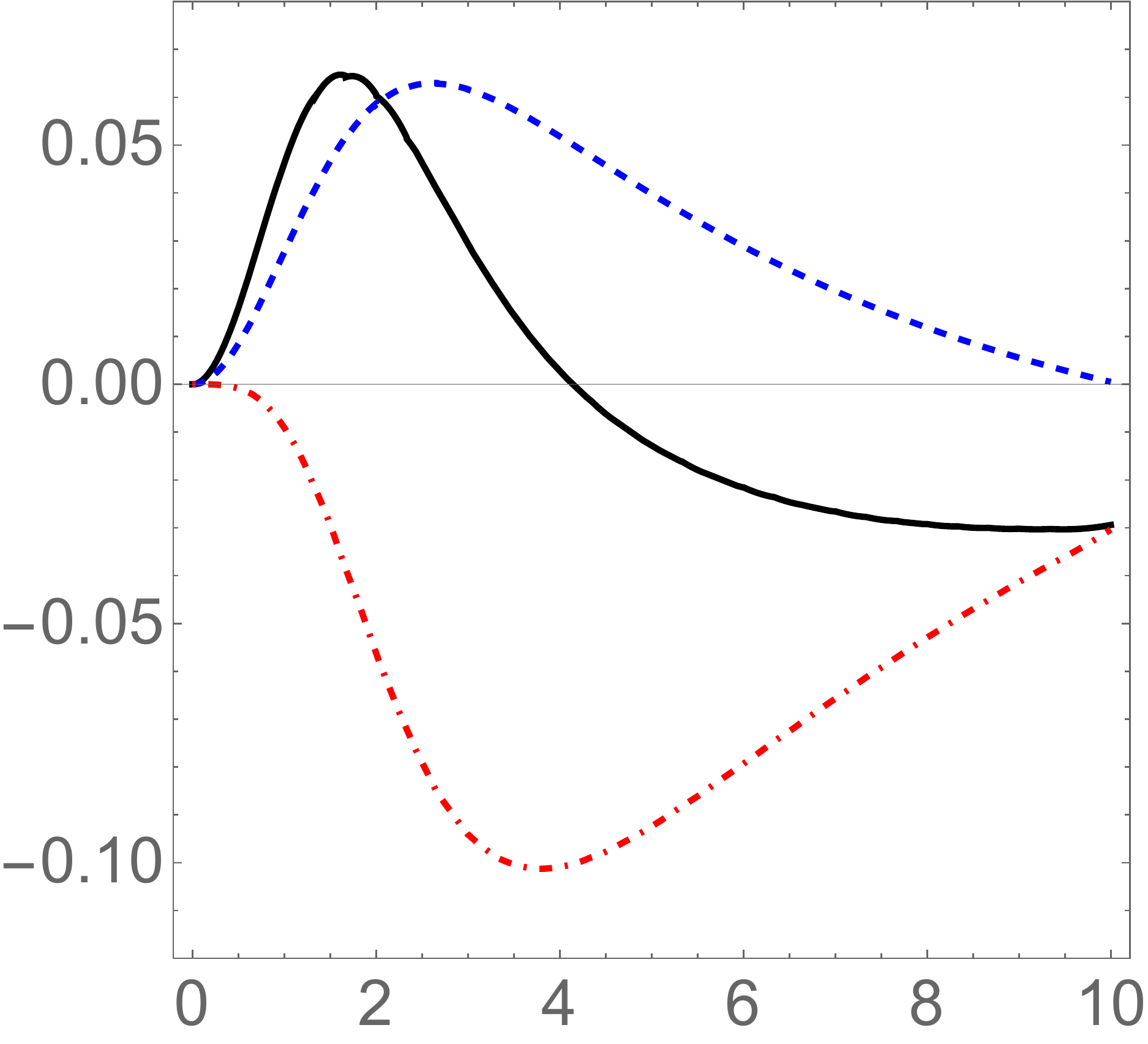}
\caption{Perturbative (upper plot) and instanton-induced (lower plot) spin-dependent potentials
 for charmonium. The black solid, blue dashed and red dash-dotted lines are  for $r^2 V_{SS}, r^2 V_{SL},
  r^2 V_{T}$, in ${\rm GeV}^{-1}$, versus $r$ in  ${\rm GeV}^{-1}$.
}
\label{fig_spin_V}
\end{center}
\end{figure}

\subsection{Lattice studies} \label{sec_lattice_SS}

 The
 static spin-spin potentials have been evaluated on the lattice, using correlators of Wilson   lines with explicit field strengths. In particular,
 Koma and Koma~\cite{hep-lat/0609078} find that while $V_{SS}$ is indeed rather short range, it does not fit to  the vector+scalar exchange paradigm:
 a pseudo-scalar glueball exchange has been proposed. Skipping  a decade, Kawanai and Sasaki 
\cite{1508.02178}  accurately derived the central and spin-spin potentials for the $\bar c c$ and $\bar c s$  families,  and 
 used them in effective Schrodinger equations for a spectroscopic analysis with nice agreement  with the spectroscopic data. 
 In Fig.~\ref{fig_Vss_C_inst} we show (solid line) their exponential fit (for $\bar c c$).
 \begin{equation} 
 V_{SS}^{\bar c c}\equiv \alpha e^{-\beta r}, \,\,\, \alpha= 2.15 \, {\rm GeV}, \,\,\,\beta=2.93 \,{\rm GeV}
 \label{eqn_VSS_lat}
\end {equation}
 Let us   try to interpret this result: Naively, the coefficient in the exponent 
 should be the mass of the object exchanged: for a gluon it should be about half of glueball masses,
 of the order of $1\, GeV$ or so. It is in fact larger by a factor of 3. Why is   it  so large?  
 
 Fortunately, one can check it phenomenologically and find that it does describe well the $2S$ to  $1S$ splitting in charmonium.  Using this $V_{SS}$  parameterization from \cite{1508.02178}  we calculate 1S triplet-singlet splitting to be $113.7\, MeV$ (exp: $113\, MeV$) and
 2S to be $47.3 ,\ MeV$ (exp: $51\, MeV$).  While not perfect, this version of $V_{SS}$ 
 clearly does the job, at least in charmonium.

\begin{figure}[h]
\begin{center}
\includegraphics[width=6cm]{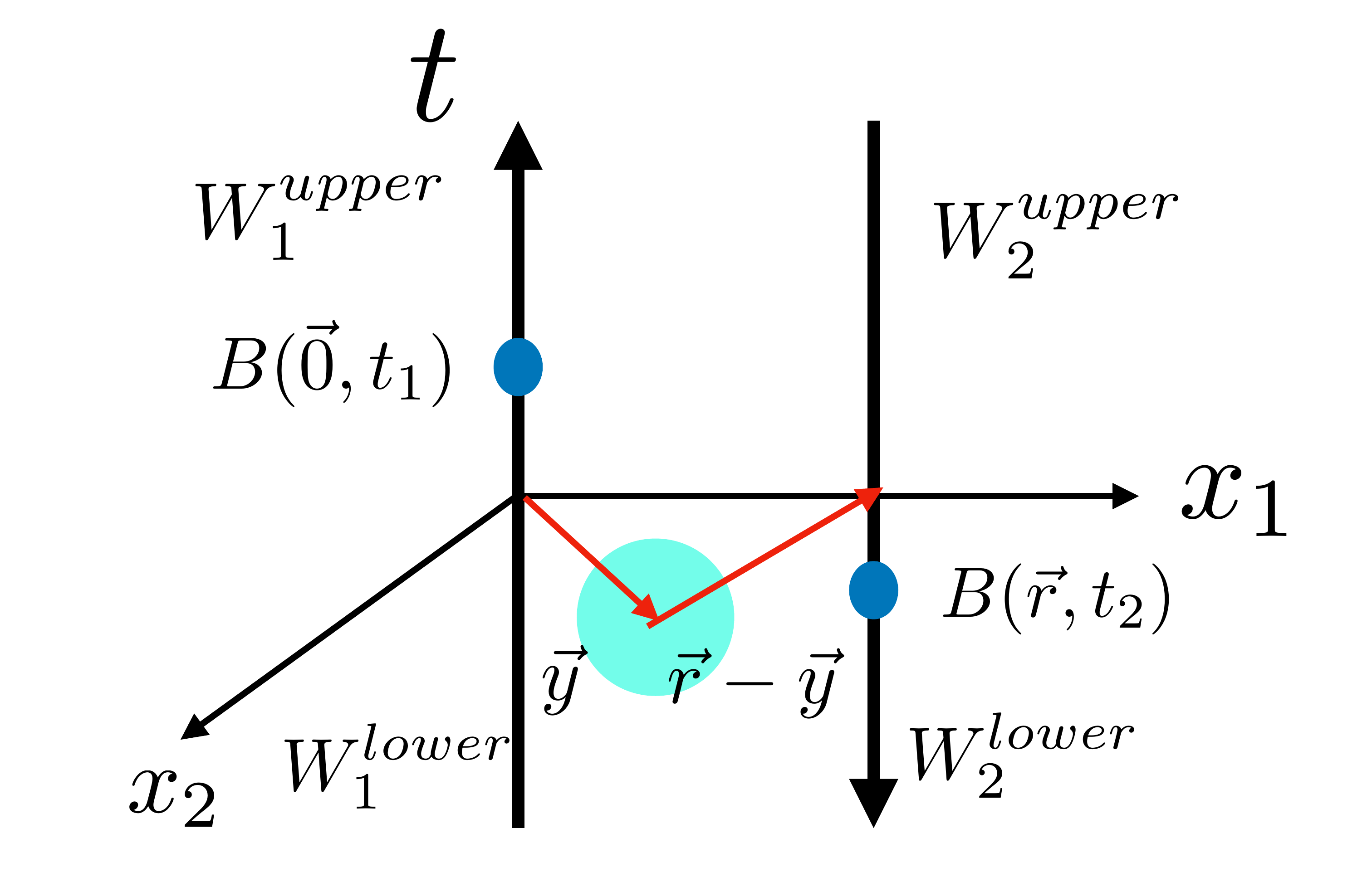}
\caption{Notations of vectors in evaluation of Wilson lines with two field strengths.
}
\label{fig_WBWB}
\end{center}
\end{figure}

\subsection{Spin-dependent forces from instantons} \label{sec_spindep_theory}
We decided first not to rely on general relations, sometimes leading to singular functions, and
evaluated the correlators of Wilson lines with fields directly.

The notations we use for the evaluation of the spin-dependent potentials are shown 
in Fig.~\ref{fig_WBWB}. The Wilson line $W_1$ is located at the origin of the spatial coordinates $\vec 0$, and
$W_2$ at $\vec r=(r,0,0)$. Two magnetic fields are inserted  on them,  and act at times $t_1$ and $t_2$,
respectively. The SU(2) color matrix associated with the field $B_m$ is $\tau^m$, and  the overall color trace 
defines the following tensor
$$ C^{m \bar m}={\rm Tr}\big[W_1^{upper} (W_2^{upper})^+ \tau^{\bar m}   (W_2^{lower })^+ W_1^{lower} \tau^{ m}\big] $$
This construction is implicitly included in what we denote by the ``double average",
for example the spin-spin potential  is related to correlator of two magnetic fields 
integrated over their time difference

\begin{equation} \label{eqn_double_average_B}
V_4(r)=\int dt \langle \langle B^{ma}(\vec 0, 0) B^{ ma}(\vec r, t)  \rangle \rangle \end{equation}
The center of the instanton is put at time zero and spatial  location $\vec y$.  On dimensional grounds, 
the instanton of size $\rho$ should contribute to those potentials $\sim  \kappa{1/\rho^3} F(r/\rho)$.

\begin{figure}[htbp]
\begin{center}
\includegraphics[width=5cm]{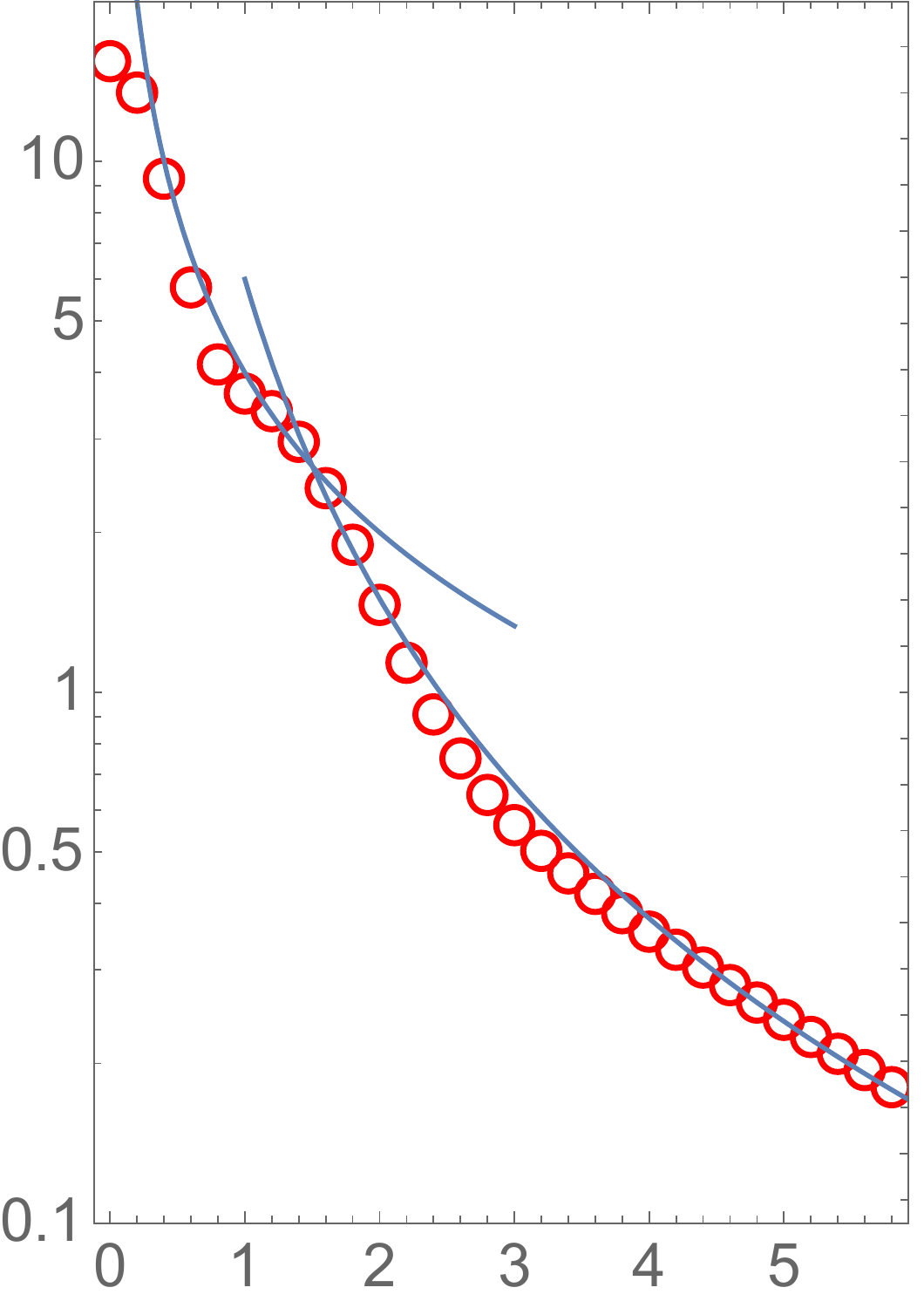}
\caption{The dependence of the integrated correlator of two magnetic fields (\ref{eqn_double_average_B})  on the distance, in units of $r/\rho$,
is shown by points. The two thin lines are  drawn for comparison, the upper is $\sim 1/r$ and the lower $\sim 1/r^2$.
}
\label{fig_BB}
\end{center}
\end{figure}

The field strength for an  instanton (in regular gauge), located at point $y$, is
\begin{equation}
G^a_{\mu\nu}(x)=-4 \eta^a_{\mu\nu} {\rho^2 \over \big((x-y)^2+\rho^2\big)^2}
\end{equation}
Four Wilson lines are written in their canonical form $W=cos(A)+i (\vec n \vec \tau) sin(A)$
with certain angles, which in regular gauge are

\begin{widetext}
\bea
&&A_1^{lower}=-\big[\pi+2 arctan({t_1 \over \sqrt{y^2+\rho^2}} )\big]{1 \over \sqrt{1+\rho^2/y^2}} \nonumber\\
&&A_2^{lower}=-\big[ \pi+2 arctan({t_2 \over \sqrt{r^2+y^2-2 r y cos(\theta) +\rho^2}}) \big] 
{1 \over \sqrt{1+\rho^2/(r^2+y^2-2 r y cos(\theta) )} }
\eea
\end{widetext}
The ``upper" lines are given by the same expressions with opposite signs of arctan. If there is no field insertions,
these terms cancel as they should, and their sum become the angles we already met in the expression for central potentials.

In total, the correlator 
includes  integration over $d^3 y,t_1,t_2$ . With  the azimuthal angle $\phi_y$ 
being irrelevant, it is 4-dimensional integral  performed numerically. 
The results are shown in Fig.\ref{fig_BB}. In this log plot one can see that the resulting spin-spin force
rapidly decreases with distance. For comparison, we show two lines with $1/r$ and $1/r^2$, which
approximately reproduces the behavior. One should not however conclude that they give correct analytic 
asymptotic: in fact, expanding the integrand in inverse powers of $r$ leads to divergences of the remaining integrals,
indicating that the true asymptotic behavior cannot be just powers.

In the instanton vacuum, with (anti) selfdual fields  $\vec E^a=\pm \vec B^a$ in Euclidean signature, all potentials in (\ref{SD}) can be tied to the central potential $V_C(r)$~\cite{Callan:1978ye,Eichten:1980mw,Turimov:2016adx}

 \begin{widetext}
\bea
V_C(r)+2V_1(r)\rightarrow 0 \qquad
 V_2(r)\rightarrow \frac 12 V_C(r) 
 \qquad V_3(r)\rightarrow -\bigg(\frac 1rV_C^\prime-V_C^{\prime\prime}\bigg)
 \eea
and the simplified spin-orbit contribution

\be
\label{SD1}
\bigg[\frac 1{2m_Q^2}\frac 1r \frac d{dr} V_C(r)\bigg]  (S_{Q}+S_{\bar Q})\cdot L_{Q}\equiv 
\bigg[\frac 1{2m_Q^2}\frac 1r \frac d{dr} V_C(r)\bigg]   S\cdot  L
\ee
\end{widetext}
 Schematically, 
the central electric interaction $\langle E_i^a(x)[x,0]^{ab}E_i^b(0)\rangle$   is amenable to the spin-orbit interaction $\langle E_i^a(x)[x,0]^{ab}B_i^b(0)\rangle$ 
and  also the spin-spin  and tensor interaction $\langle B_i^a(x)[x,0]^{ab}B_i^b(0)\rangle$.

\begin{figure}[b!]
\begin{center}
\includegraphics[width=6cm]{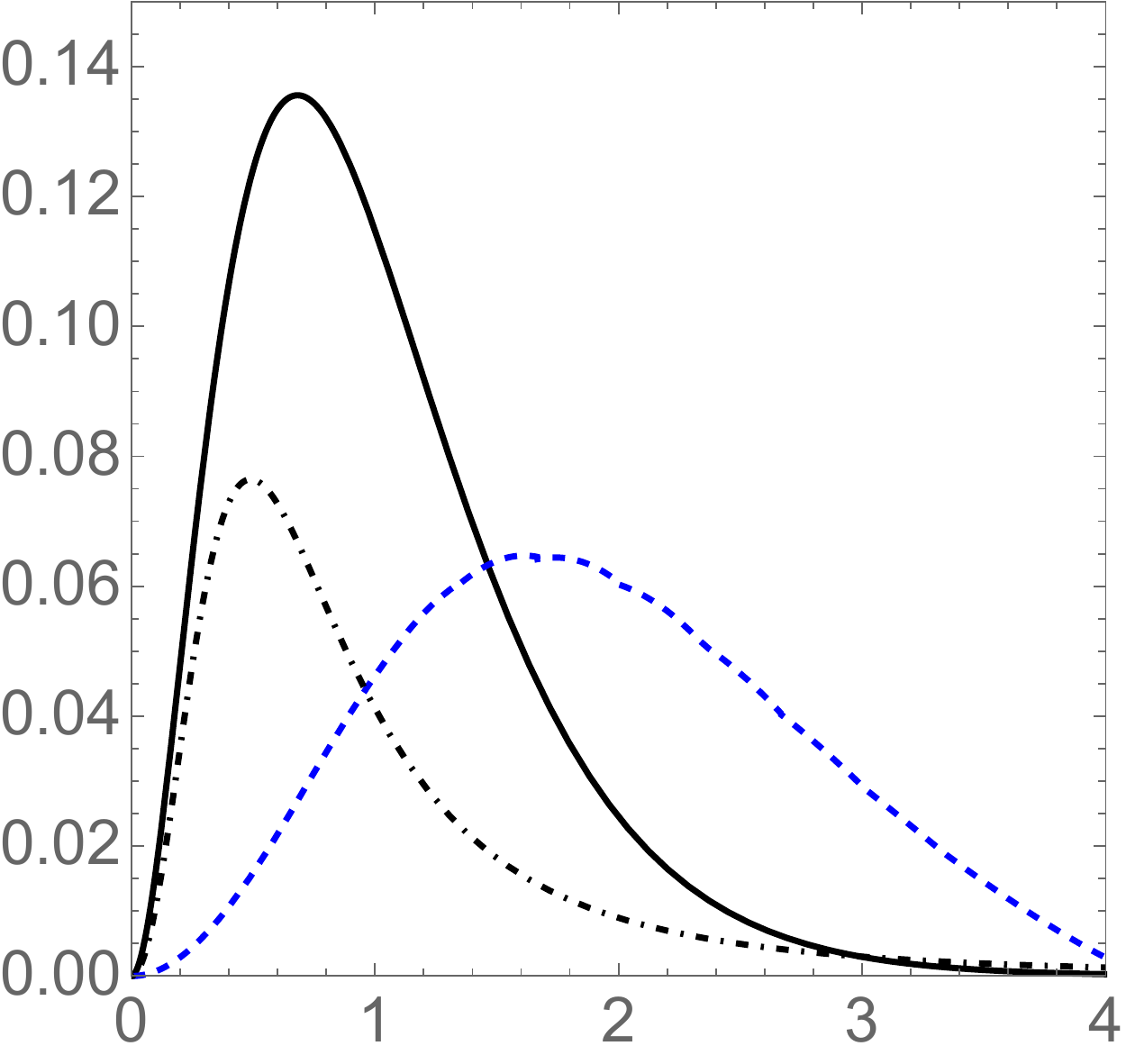}
\caption{Spin-spin potential  for $\bar c c$ system multiplied by distance squared $r^2 V_{SS}\, (GeV^{-1})$ versus $r \, (GeV)^{-1}$.The solid line is the exponential fit (\ref{eqn_VSS_lat})  to lattice measurements \cite{1508.02178}. The dash-dotted line shows the regulated laplacian of the Coulomb potential (\ref{regulated_Laplacian}) with $\delta=0.6\, GeV^{-1}$. The blue dashed line show the instanton contribution. Note that the area under the last two curves is roughly equal to that under solid one,
the lattice fit.
}
\label{fig_Vss_C_inst}
\end{center}
\end{figure}

 In Fig.\ref{fig_Vss_C_inst} we compare the perturbative (dashed-dotted-black) and nonperturbative (dashed-blue)
 spin-spin potentials with the lattice result  (solid-black). Recall that  the Laplacian of the Coulomb force is a
  delta function, well known in atomic and positronium physics. However, one should use Coulomb potential $1/r$
   in some regulated form, for several reasons.
First, when the potential gets as deep as (minus)  mass, the nonrelativistic approximation itself
  should break down, so the spread cannot be smaller than $O(1/m_Q)$.  
  Second, lattice studies include non-locality in the form of the lattice spacing $a$.
Third, a Gaussian-smeared delta function is actually used in many spectroscopic calculations, see e.g. \cite{hep-ph/0505002}.  
For the perturbative spin-spin interaction, we use a simpler version,
as  the  Laplacian of a Coulomb contribution regulated by a fixed parameter 
$\delta=0.6\, {\rm GeV}^{-1}$
\begin{equation} V_{SS}^C(r)= -{A \over 3 m_Q m_{\bar Q}}  \vec \nabla^2 {1\over \sqrt{r^2+\delta^2}}
\label{regulated_Laplacian}
\end{equation}
For the nonperturbative  spin-spin  part, we use the instanton contribution, which 
is given by the Laplacian  acting on the central potential
\begin{equation} 
V_{SS}^{\rm inst}(r)= {1 \over 3 m_Qm_{\bar Q}}  \vec \nabla^2 V_C^{\rm inst}(r)
\end{equation}

All the potentials shown are multiplied by $r^2$, as they appear in the volume integrals. So, if the wave functions
are  roughly constant at this small scale, the area under the curves in Fig.~\ref{fig_Vss_C_inst}
gives an  estimate  of  the relative contributions. So, one finds  that the perturbative and non-perturbative
parts contribute $comparably$ to the total spin-spin potential, in sum roughly reproducing the lattice-based
result. This observation is 
among the main findings of this work.

Of course, the actual magnitude of the spin splittings depends on the quark masses involved, explicitly via
$1/m_Q^2$ factors and implicitly, via the wave functions. Lighter quarks lead generally to hadrons
of larger size, or smaller wave functions at small distances.

 The matrix elements of all three potentials shown in  Fig.~\ref{fig_Vss_C_inst}
were evaluated using wavefunctions from the Cornell potential irrespective of flavor. The results for the spin-spin 
contributions are in Table~\ref{tab_1S}. The results show  that   perturbative and non-perturbative 
contributions for heavy quarkonia do  reproduce the splittings if {\em taken together}.
Already for heavy-light mesons one can see that some additional effect seems to be missing:
it will indeed so as will be explained below.
Furthermore, for the mesons containing only light quarks both contributions together definitely
 fail to describe large splittings seen experimentally.  Clearly some other effects are in play here, 
 to be discussed below.

We continue our phenomenological description of the spin-dependent interactions, by
considering  the next four states in the 1P shell, with the onset of the spin-orbit  and tensor potentials
The empirical information to be used on several mesonic families is listed in  \ref{tab_1P}.
The theoretical estimates follow from  1P wave functions derived with   standard Cornell potential. 
For both components of the central potential  -- perturbative and instanton-induced -- we calculated the matrix elements of the three spin-dependent potentials using
the formulae given above, and then calculated their matrix elements in  the 1P shell.

One observation from this table  is that  the ``theory" (the sum of the last two rows) and  the lattice-induced one (the second row)
are basically in agreement.  The other is that their disagreement with experiment (the upper raw)
 is dramatically growing
for light quark systems. This implies that for light quarks something important 
is missed, in the description as developed so far.

 \begin{widetext}
\begin{table}[t] 
\caption{``Hyperfine" splittings of certain $L=0$ mesons with $J=1$ and $J=0$. The first row of numbers 
shows the experimental values (MeV) (rounded to 1  MeV).  The second row gives the matrix elements
of the lattice-based spin-spin potential (\ref{eqn_VSS_lat}), the next two rows are the (regulated) Coulomb
and instanton-induced spin-spin contributions.}
\begin{center}
\begin{tabular}{|c|c|c|c|c|c|}
\hline
flavors & $M_{\Upsilon}-M_{\eta_b}$ & $M_{J/\psi}-M_{\eta_c}$ & $M(D^*)-M(D)$ & $M(K^*)-M(K)$ & $M(\rho)-M(\pi)$ \\
\hline
Exp &                                                                     61. & 116.  & 137. & 398. & 636. \\
$\langle V_{SS}^{\rm lat}/3m_Qm_{\bar Q} \rangle $      & 46. & 108.  &  98. &  170.& \\
$\langle\vec \nabla^2 V_C/3m_Qm_{\bar Q}  \rangle$   & 28. &   58.  & 48. &    82.&\\
$\langle\vec \nabla^2 V_{\rm inst}/3m_Qm_{\bar Q} \rangle $ & 7.&  30.  & 48. &   90.& \\
\hline
\end{tabular}
\end{center}
\label{tab_1S}
\end{table}%

\begin{table}[t]
\caption{Groups of three  matrix elements of spin-spin, spin-orbit and tensor  potentials $V_{SS}, V_{SL}, V_T$, respectively. The first one in each group is  the observed ``exp" value from Table \ref{tab_abcX}, the second and the third values are the perturbative and instanton-induced contributions, corresponding to the Coulomb 
and instanton-induced  parts of the central potential. 
}
\begin{center}
\begin{tabular}{|c|c|c|c|c|c|c|c|c|c|}
\hline
             &   SS ``exp" & SS pert & SS inst & SL ``exp" & SL pert & SL inst &  T ``exp" & T pert & T inst\\
\hline
$\bar{c} c$  & 0.1 &   0.56    &    1.9   &  35.        & 3.2 &    3.8                       & 10.  & 5.8   & -5.7\\
$\bar{s} q$  & 151. & 5. &  29. & 6.7 & 31. & 38.  & -5.6 & 58. & -46. \\
$\bar{q} q$  &  68. & 8. &  48. & 0.5 & 52. & 64.  &  -36. & 96. &   -78. \\
\hline
\end{tabular}
\end{center}
\label{tab_1P}
\end{table}%
\end{widetext}

\subsection{Contributions of $I\bar I$ molecules to spin forces} \label{spin_molecules}
Like in subsection \ref{central_molecules}, let us start with the $I\bar I$ molecules oriented in the time direction, for which $\vec E\sim G_{4m}$ and $ A_4$ are time-odd. Therefore, as we noted before,
there is no contribution to the central potential.
 
 Yet the
 {\em spin-spin and tensor} forces are not at all zero, as  the main contributors  to the correlator come from the $magnetic$ fields
$\vec B$, which  have {\em the same sign} in the  $I$ and $\bar I$ parts of the molecule! 

A detailed description of the $I\bar I$ configuration can be achieved  by the Yung ansatz, which is quite accurate. The corresponding expressions
for field strengths  were obtained in Mathematica, which  are  way too  complicated to be given here. 
Yet the field strengths are approximately additive, and this approximation will be used in what follows.

Let us start with the correlator of two magnetic fields,   measured at integrated times $t_1$ and $t_2$ 
on  two  temporal lines separated by a spatial distance $r$ 

\begin{widetext}
\bea
 \label{eqn_molec_t}
\int dt\,\langle B^{ma}(0,\vec 0) B^{ma}(t, \vec 0)\rangle =&&\int dt_1 dt_2 d^3y (48 n_{mol} \rho^4)\nonumber\\
&&\times \bigg({1 \over ((t_1 - R/2)^2 + \rho2 + y^2)^2} 
+ {1\over ((t_1 + R/2)^2 + \rho2 + y^2)^2}\bigg) \nonumber\\
  && \times 
   \bigg({1 \over ((t_2 - R/2)^2 + \rho2 +
        y_2^2)^2} +{ 1\over ((t_2 + R/2)^2 + \rho2 + y_2^2)^2}\bigg)
        \eea
        \end{widetext}
Here  $y^2$ and $y_2^2=y^2+r^2- 2 r y cos(\theta_{ry})$ are the squared 3d distances
to the first and second Wilson lines.  We assumed additivity of the magnetic fields from
$I$ and $\bar I$,  and 
  ignored the color rotation angles on  the Wilson lines
  (which vanish at the instanton center, as we already know).
  
 At zero distance between the line $r=0$, all integrations can be done analytically,
with the result

\be  
\label{SSMOL}
\int dt \,\langle B^{ma}_1(0, \vec 0) B^{ma}_2(t, \vec 0) \rangle_{r=0} = 12 \pi^4 \rho n_{mol} 
\ee
For qualitative orientation, one can compare this expression to (the extrapolation of) the lattice
spin-spin potential (\ref{eqn_VSS_lat}) to $r=0$: they are  equal if $n_{mol} \approx 7 \, fm^{-4}$. 

Note that in this approximation there is no dependence on the (time)  separation $R$ of $I$ and $\bar I$ centers. 
For nonzero $r$ the volume integral over $y, cos(\theta_{ry})$ is done numerically, see the
results in Fig.~\ref{fig_molec_corr}.

Now let us calculate the same correlator, but for a molecule rotated so that its 
vector $R_\mu$ joining  the centers, lines in the spatial direction, say the same as the  vector $\vec r$ between the Wilson lines. 
The electric and certain magnetic fields exchange places and the expression changes into 

\begin{widetext}
\bea \label{eqn_molec_x}
\int dt\,\langle  B^{am}(0, \vec 0) \vec B^{am}(t, \vec 0) \rangle &=&\int dt_1 dt_2 dy_1 2\pi y_\perp  dy_\perp (48 n_{mol} \rho^4)\nonumber\\
&&\times \bigg({1 \over (t_1^2 + \rho^2 + (y_1-R/2)^2+y_\perp^2)^2}
 - {1\over (t_1^2 + (y_1+R/2)^2+ \rho^2 + y_\perp^2)^2}\bigg) \nonumber\\
 && \times \bigg({1 \over (t_2^2 + \rho^2 + (y_1-R/2+r)^2+y_\perp^2)^2} - 
   {1\over (t_2^2 + (y_1+R/2+r)^2+ \rho^2 + y_\perp^2)^2}  \bigg)\nonumber\\
  \eea
\end{widetext}
In Fig.~\ref{fig_molec_corr} we compare the correlators for two molecule orientations with averaging factors, 
$(1/4) (\ref{eqn_molec_t})$ and $(3/4) (\ref{eqn_molec_x})$ by upper and lower points, as well as their sum (the line). 
So, while the two contributions have different dependence on $r$, the orientation-average
$V_{SS}(r)$  decreases monotonously, and has the effective width (at half maximum) close to the average instanton size $\rho$. 
Note that this range is shorter than $V_{SS}(r)$  from the instantons alone (calculated from the Laplacian of the contribution to the central potential
in the previous subsection), and it does not have a negative part. It is therefore closer in shape to what was found on the lattice.

\begin{figure}[htbp]
\begin{center}
\includegraphics[width=8cm]{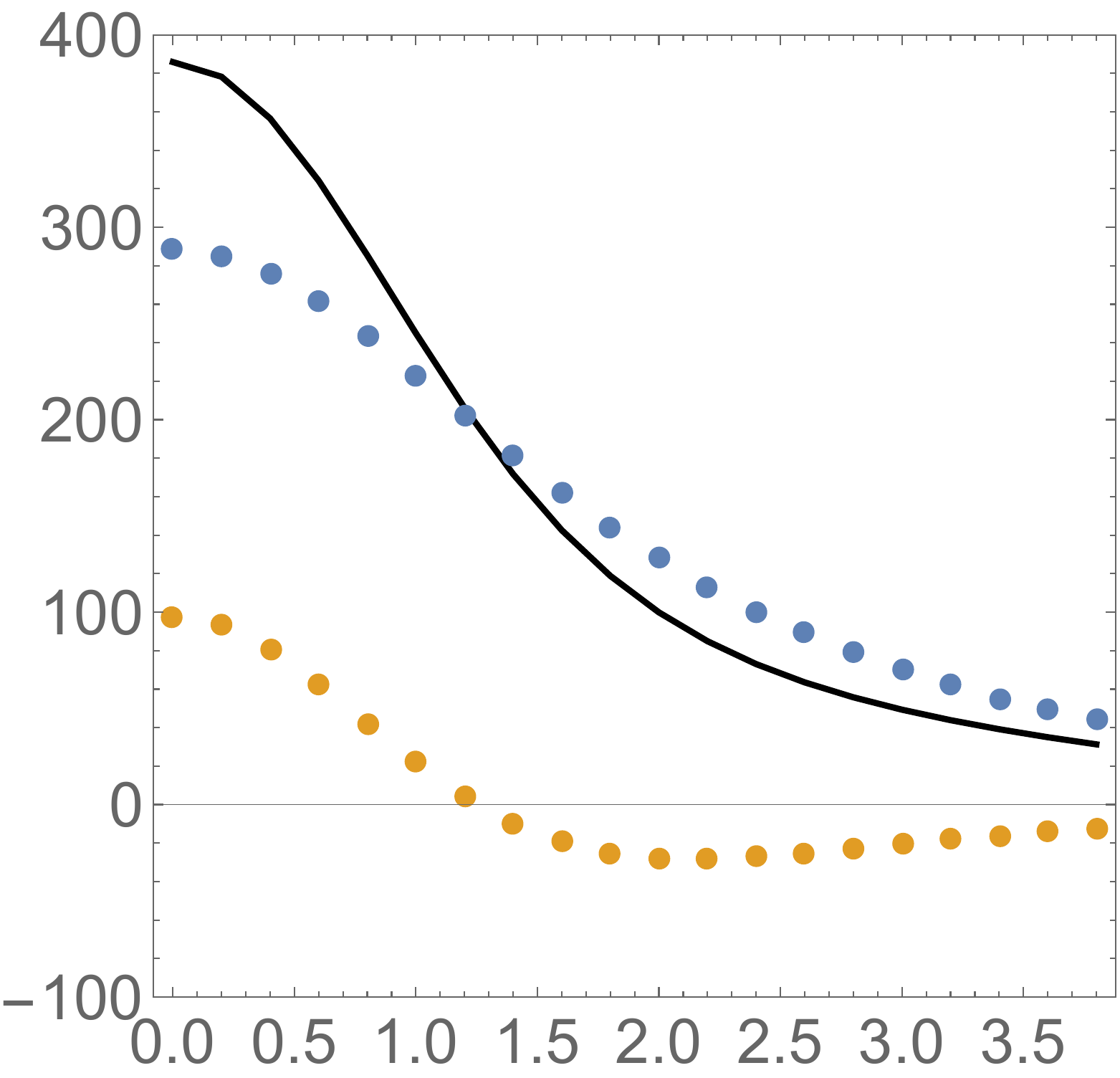}
\caption{The normalized correlators $(1/4) (\ref{eqn_molec_t})$ (upper points)
and $(3/4) (\ref{eqn_molec_x})$  for $R=\rho$ (lower points) in units of $n_{mol} \rho$, versus the
distance between the Wilson lines $r/\rho$. The  solid line shows their sum. }
\label{fig_molec_corr}
\end{center}
\end{figure}

The {\em spin-orbit} force is qualitatively different from spin-spin ones, being related to a correlator $ \int dt \vec E \vec B$. 
For  a time-oriented molecule $R_\mu=(0,0,0,R)$ $\vec E(t)$ is time-odd and $\vec B(t)$ is time-even, so for this molecule orientation
this correlator vanishes. 
The situation is different for  a space-oriented molecule, say along $x_1$. Now  $G_{1m}, m=2,3,4$ 
 fields are odd in $x_1$ inversion, but everything is even along the time axes. The expression for the correlator
 can easily be constructed and calculated in a way similar to what was done above for $V_{SS}(r)$

 We conclude that the  ``molecule-induced" spin forces are  quite similar to the  instanton induced ones
 (apart of course normalization to quite different densities) at small distances, but are {\em significantly reduced}
 at  $r> 2\rho$ due to the partial cancellations by different orientations. 
 
 In the previous section we have seen that, in the approximation of molecules represented as two independent pseudoparticles,
the non-perturbative  $V_{SS(r)}$ have about the right magnitude, but somewhat too large range. We now see that a
better description of the field structure has the potential to remedy this problem.

  \subsection{Contributions of the fixed energy tunneling configurations  to spin forces} \label{fixedE}
  
  The finite energy tunneling configurations contribute to the spin forces, through close Wilson loops
dressed by $E,B$ fields. However, since these configurations cease to be self-dual 
for $0\leq k<1$, the induced spin potentials are in general independent. Moreover, these
configurations carry  t-odd electric fields,  and are  purely magnetic at the exit points ($t_E=0$).
They may only contribute to the spin-spin and tensor potentials.

The explicit expression of the magnetic field in Euclidean time is 

\begin{widetext}
\bea
B^{ma}(t_E,\vec x)&=&\delta^{ma}\bigg(-\frac{16\rho^2}{\mathbb D}f+\frac{8\rho^2r^2}{\mathbb D^2}fF+\frac{32\rho^3t_Er^2}{\mathbb D^2}f^\prime
+\frac{32r^2\rho^2}{\mathbb D^2}f^2\bar F\bigg)\nonumber\\
&+&x^mx^a\bigg(-\frac{8\rho^2}{\mathbb D^2}fF-\frac{32 t_E\rho^3}{\mathbb D^2}f^\prime
+\frac{32\rho^2}{\mathbb D^2}f^2F\bigg)\nonumber\\
&+&\epsilon^{mai}x^i\bigg(\frac{16\rho}{\mathbb D}f +\frac{4\rho}{\mathbb D^2}fF\bar F+\frac{16t_E\rho^2}{\mathbb D^2}f^\prime\bar F
-\frac{32\rho^3}{\mathbb D^2}(2r^2+\bar F)f^2\bigg)
\eea
\end{widetext}
with $f^\prime=\partial_\xi f_{k}(\xi)$, 
\bea
\label{DTR}
\mathbb D=(t_E^2+r^2+\rho^2)^2-4t_E^2\rho^2
\eea
and 
\bea
F=t_E^2+r^2+\rho^2\qquad \bar F=-t_E^2-r^2+\rho^2
\eea
The magnetic field in Minkowski signature follows from the substitution $t_E\rightarrow -it$. 

The induced spin-spin interaction follows from the complexified path in the t-plane shown in Fig.~\ref{fig_zig}.
 An estimate of the 
magnitude of the spin-spin interaction,  follows from zero separation between a path ${\cal C}$ and $\bar {\cal C}$

\begin{widetext}
\bea
\label{DTRX}
\int_{\cal C} dt\int_{\bar{\cal C}} d\bar t\, \left<B^{ma}(-it, \vec 0)B^{ma}(-i\bar t, \vec 0)\right>
=[3\,2^8\rho^4n_{k}]\int_{\cal C} dt\int_{\bar{\cal C}} d\bar t\, \int d^4Z\bigg(\frac{f\bar f}{\mathbb D\overline{\mathbb D}}\bigg)
\eea
\end{widetext}
with now

\bea
\label{DTRX1}
&&\mathbb D\rightarrow ((-it-Z_4)^2+{\vec Z}^2+\rho^2)^2-4(-it-Z_4)^2\rho^2\nonumber\\
&&\overline{\mathbb D}\rightarrow ((-i\bar t-Z_4)^2+{\vec Z}^2+\rho^2)^2-4(-i\bar t-Z_4)^2\rho^2\nonumber\\
\eea
and $F=f_k(\xi)$ and $\bar f=f_k(\bar \xi)$ with

\bea
&&\xi\rightarrow \frac 12{\rm ln}\bigg(\frac{(-it-Z_4+\rho)^2+{\vec Z}^2}{(-it-Z_4-\rho)^2+{\vec Z}^2}\bigg)\nonumber\\
&&\bar\xi\rightarrow \frac 12{\rm ln}\bigg(\frac{(-i\bar t-Z_4+\rho)^2+{\vec Z}^2}{(-i\bar t-Z_4-\rho)^2+{\vec Z}^2}\bigg)\nonumber\\
\eea
with the Euclidean-Minkowski time assignments in (\ref{CPATH}).

At the sphaleron point $f_{k=0}=1/2$, and
 (\ref{DTRX}) reduces to

\begin{widetext}
\bea
\label{DTRX1}
&&\int_{\cal C} dt\int_{\bar{\cal C}} d\bar t\,\left<B^{ma}(-it, \vec 0)B^{ma}(-i\bar t, \vec 0)\right>\rightarrow 
[192\rho^4n_{k=0}]\int_{\cal C} dt\int_{\bar{\cal C}} d\bar t\, \int d^4Z \nonumber\\
&&\times 
\frac 1{(((-it-Z_4)^2+{\vec Z}^2+\rho^2)^2-4(-it-Z_4)^2\rho^2)(((-i\bar t-Z_4)^2+{\vec Z}^2+\rho^2)^2-4(-i\bar t-Z_4)^2\rho^2)}\nonumber\\
\eea

Overall time translational
 invariance allows us to set  $Z_4=0$ in the integrand, with the final result 

\bea
\label{DTRXX}
\frac{\int_{\cal C} dt\int_{\bar{\cal C}} d\bar t\, \left<B^{ma}(t, \vec 0)B^{ma}(\bar t, \vec 0)\right>}{\int dZ_4}=\frac{\pi^4}{4\rho^3}
\eea
\end{widetext}
This result follows  by deforming the complexified path to the real axis, without encountering poles for ${\cal T}_{k=0}<\rho$.
The spin-spin interaction (\ref{DTRXX}) stemming from the finite energy tunneling configurations, is comparable
to the spin-spin interaction (\ref{SSMOL}) from the  thimbles. At the sphaleron point and for zero separation,
they are comparable when

\bea
\frac{n_{k=0}}{n_{\rm mol}}=\frac{12\pi^4}{192\pi^4/4}=\frac 14
\eea

\section{Interquark forces induced by instanton zero modes \\ ('t Hooft effective Lagrangian)} \label{sec_Hooft}
\subsection{Pseudoscalar mesons which are and are not the Nambu-Goldstone modes}\label{sec_fixing_G}

It is known since 1970's that in the chiral limit  light-quark pseudoscalar mesons consist
of octet $\pi,K,\eta$ Nambu-Goldstone modes and a singlet $\eta'$ which does $not$ belong
to that set, and is instead very heavy.  It happens because  the   t' Hooft effective Lagrangian
violates $U_A(1)$ and  quite effectively generates flavor-changing annihilation processes
$$ \bar u u \leftrightarrow \bar d d  \leftrightarrow \bar s s $$
When a strange quark mass is added, one finds $\eta-\eta'$ mixing which
describes well the phenomenology of their decays. It  sharply contrasts with the vector counterparts,
$\rho,\omega,\phi$ mesons, for which the mixing is very small.

To avoid these annihilation processes, one can discuss ``flavored"
channels, like $\pi^+=\bar d u$, in which case the  t' Hooft effective Lagrangian acts as a force between the quark and antiquark.  Yet it is still different from
the conventional forces we so far discussed, in that it exists for quarks of {\em different flavors} only. 
In Fig.\ref{fig:pi_rho} we illustrate how
the  t' Hooft induced vertex  operates in the  pion and rho charged meson channels.

Before we turn to specifics, let us comment on some lattice implementations of theories with
different quarks. The simplest case is a single quark flavor theory, $N_f=1$. In it the 't Hooft action is a 2-fermion
operator: it gives the quark an effective mass, but does not generate annihilations or interactions as we noted. 
The only pseudoscalar mesons is the analog of $\eta'$, no $\pi,K,\eta$ modes exist.

In lattice implementations nowadays one uses gauge ensemble for physical QCD, but still use
certain diagram and mass selections aiming to study certain unphysical particles. For example, 
in \cite{Koponen:2017fvm},  an artificial particle called $\eta_s$  made of $\bar s s$ was intoduced, with no  mixing
to  other flavors. Its properties are deduced from the $connected$ diagram, without the disconnected 
one in which the flavor is changed. One may explain its absence by introducing an $extra$  valence quark species $s'$, 
and view source/sink operators as flavor-nondiagonal $\bar s s'$. Our comment is that such setting, while
eliminating the annihilations, still keeps the  't Hooft effective Lagrangian producing a force between the quark and antiquark, as they are not of the same flavor. 
Apart from the mass values, this channel is no different from say $\pi^+=\bar d u$ we consider.

In our first take on spin-spin forces in (\ref{eqn_SS_splittings}) we already observed anomalously large
splitting between the $\rho, S=1$ and $\pi, S=0$ states in the lowest light quark shell, and have seen that
the magnitude of $V_{SS}(r)$ discussed so far, is not sufficient to explain it. 

(In particular, adding
$V_{SS}(r)$ as detailed above to the basic quark model, with the Cornell potential and  a light quark constituent mass of
$0.35\, {\rm GeV}$, reduces the ground state 1S mass  substantially, from $1.4 \, {\rm GeV}$ to  about $0.85\, {\rm GeV}$. 
Yet it is still far from the observed pion mass of $0.138 \, {\rm GeV}$ and from zero, which it should be in the chiral limit.)
Of course, a correct pion mass  can only be explained  in an approach, consistently explaining chiral symmetry breaking, as well as other parameters of chiral perturbation theory. Two famous models --
NJL and ILM -- are well known examples of that.

Before we focus on the  theory, let us  return for a brief moment to the  phenomenological
constituent quark model, to see what kind of interaction one needs  to add in order to obtain the experimentally observed pions. Let us assume that  there is an additional interaction, that operates  {\em only} between quarks of different flavors,
on top of the  flavor-blind forces so far discussed. Note that the usual discussion of $\pi-\eta'$ splitting is
based on ``annihilation channels" (like $\bar u u \leftrightarrow \bar d d$) as we continue to consider only charged
mesons, like $\bar d u$. 

The 't Hooft Lagrangian is usually written as a local operator {\em a la} NJL one, corresponding to 
a potential of the type $G_{\rm Hooft} \delta^3(\vec r)$ potential.  This form assumes  that  the
instanton size $\rho$ is much smaller than the hadronic sizes, as it appears
in the standard LSZ ``amputation" of external propagator lines. Unfortunately, in the Euclidean setting
of the instanton calculus, taking these lines on-shell amounts to the limit $p^\mu \rightarrow 0$  for all components.
And yet, it is clear that the process is nonlocal, and that the scale of its size is $\rho$. Thus in model applications
one include form-factors, usually the Fourier transform of the zero modes. 
We will use another  ``regulated delta function" form described in section \ref{sec_nonlocality_of_Hooft}.

(Comment:
When the  nonlocality induced by the parameter $\rho$ is small compared to the hadron size, it can be well approximated by a delta function. However, this is not true for pions. This can be
seen from  the pion wave function  (with or without the effective 't Hooft term): 
 its $r^2 \psi^2_\pi(r)$ peaks at $r\approx \rho$. In this case. Therefore, in the pion case there  a remains certain dependence on a choice of the non-local approximation.)
 
 Finally,   the coupling strength for the t' Hooft effective interaction to add to the discussed potentials above needed
 to reproduce the physical pion mass is found to be
\begin{equation} G_{\rm Hooft}= 17 \, {\rm GeV}^{-2}  \label{eqn_G_pion}
\end{equation}

\subsection{Chiral anomaly and instanton zero modes}

It was well understood already in 1970's that instantons provide an example of
explicit violation of the $U_A(1)$ chiral symmetry. The divergence of the corresponding axial current is
proportional to the topological charge density $G\tilde G$, which in turn is a divergence of the topological Chern-Simons current. Thus the axial charge of fermions, or the number of right minus left-polarized quarks, 
is intimatly related to gauge topology.

How this happens in the instanton case was explained by t' Hooft \cite{tHooft:1976snw},  who found the
 existence of 4-dimensional bound state, or the {\em fermion zero mode} of the Dirac operator in the
 instanton background. 
Formally it amounts  to   an additional term in the 4-dimensional quark propagator
\begin{equation}  S(x,y)=\sum_\lambda {\psi_\lambda(x) \psi_\lambda(y)^+  \over \lambda+im} \end{equation}
 with the zero mode  $\lambda=0$.
The physical meaning is that  the negative energy Dirac sea creates a new state, into which the
original quark is pushed into. At the same time, a quark of {\em opposite chirality} emerges from the  Dirac sea
as a physical (positive energy) state. The infinite hotel story.

This process can be described by a ``chirality flip" of a quark. 
However, the phenomenon is not reduced just to this: $all$ light quark flavors must  experience this together
and flip their chiralities $simultaneously$.  Because of the Pauli principle, only the zero modes with different flavors can undergo 
simultaneous tunneling,  the
t' Hooft effective Lagrangian is therefore a six-quark operator, with $\bar u u \bar d d \bar s s$ quarks.
%
As emphasized by t ' Hooft, it is repulsive in the $\eta'$ channel, and solves the famous  the famous $U_A(1)$  problem. But the fact that
it is $attractive$ in the scalar $\sigma$ and pseudoscalar $\pi,K,\eta$ octet channels leads to 
even more important consequences,  breaking the $SU(N_f)$ 
 chiral symmetry spontaneously~\cite{Shuryak:1982hk}. The account for these effects led to the statistical studies of 
instanton ensembles in  the 1990's, for a review see \cite{Schafer:1996wv}. Many more studies based on 
these observations, were carried in the last two decades as well: let us just mention our recent study of the so called instanton-sphaleron 
production process,  which can be studied experimentally in colliders \cite{Shuryak:2021iqu}. 

   For  clarity, let us now discuss some technical issues already discussed
   in  the literature. The first is the explicit analytic form of the 't Hooft operator,
   which by Fierz transformations, can take several different forms.
   A straightforward
   reduction of  the product of six unitary color rotation matrices, averaged over the $SU(3)$ group,
   leads to color structures containing $f^{abc} \lambda^a  \lambda^b \lambda^c$  and $d^{abc} \lambda^a  \lambda^b \lambda^c$ terms, which is difficult to use in practice. 
   However they (and in fact all color matrices) can be eliminated using specific properties of zero modes
    producing the following result~\cite{Chernyshev:1995gj} 

\begin{widetext}
\begin{align} 
	\label{DET5}
	&{\cal V}^{L+R}_{qqq}=\frac {G_{Hooft}}{N_c(N_c^2-1)}\bigg[
	\bigg(\frac{2N_c+1}{2(N_c+2)}\bigg)
	\begin{Vmatrix}
		\overline u_Ru_L& \overline u_Rd_L & \overline u_Rs_L\\
		\overline d_Ru_L & \overline d_Rd_L & \overline d_Rs_L\\
		\overline s_Ru_L & \overline s_Rd_L & \overline s_Rs_L\\
	\end{Vmatrix}
	\nonumber\\
	&-\frac 1{2(N_c+1)}\sum_{a=1}^3
	\bigg(
	\begin{Vmatrix}
		\overline u_R\sigma^au_L& \overline u_R\sigma^ad_L & \overline u_Rs_L\\
		\overline d_R\sigma^au_L & \overline d_R\sigma^ad_L & \overline d_Rs_L\\
		\overline s_Ru_L & \overline s_Rd_L & \overline s_Rs_L\\
	\end{Vmatrix}
	+\begin{Vmatrix}
		\overline u_R\sigma^au_L& \overline u_Rd_L & \overline u_R\sigma^as_L\\
		\overline d_Ru_L & \overline d_Rd_L & \overline d_Rs_L\\
		\overline s_R\sigma^au_L & \overline s_Rd_L & \overline s_R\sigma^as_L\\
	\end{Vmatrix}
	+\begin{Vmatrix}
		\overline u_Ru_L& \overline u_Rd_L & \overline u_Rs_L\\
		\overline d_Ru_L & \overline d_R\sigma^ad_L & \overline d_R\sigma^as_L\\
		\overline s_Ru_L & \overline s_R\sigma^ad_L & \overline s_R\sigma^as_L\\
	\end{Vmatrix}\bigg)\bigg]
	\nonumber\\&+(L\leftrightarrow R)
\end{align}
\end{widetext}

with  the strength of the 6-q operator
\be
\label{22X}
 G_{\rm Hooft}={n_{I+\bar I} \over 2} \bigg(4\pi^2\rho^3 \bigg)^3 \bigg({1 \over m^*_u\rho}\bigg) \bigg({1 \over m^*_d\rho}\bigg) \bigg({1 \over m^*_s\rho}\bigg)
\ee
with the effective quark masses $m^*_q$ to be explained a bit later.

In (\ref{DET5}) all quarks and antiquarks have opposite chiralities. In  a completely massless theory  we cannot 
have ``loop" quarks by a propagator conserving chirality. However, the spontaneous 
 breaking of chiral symmetry in the QCD vacuum leads to nonzero quark condensates $\langle \bar q_f q_f \rangle \neq 0$. 
 This fact leads to ``dimensional reduction" of the six-quark operator to four (and even two-) quark expressions,
by ``looping" some quark-antiquark pairs into the corresponding condensates. For example, 
strange quark loops generate  an effective
 $ud$-interaction. It  also has a structure of two-by-two determinants, which (with some abuse of notations) we write as
\begin{widetext}
\bea
\label{24X}
{\cal V}^{L+R}_{qq}=
{\kappa}_2\,A_{2N}\,
\bigg({\rm det}(UD)+B_{2N}\,{\rm det}(U_{\mu\nu}D_{\mu\nu})\bigg)
+(L\leftrightarrow R)
\eea
\end{widetext}
with
\bea
\label{25X}
A_{2N}=\frac{(2N_c-1)}{2N_c(N_c^2-1)}\qquad B_{2N}=\frac 1{4(2N_c-1)}\nonumber\\
\eea
The  coefficient 
$$\kappa_2=3!\,G_{\rm Hooft}\langle \overline s_R s_L\rangle=3\,G_{\rm Hooft}\langle \overline s s\rangle<0\,\,,$$  is negative and thus attractive.
In the Weyl basis, 
$\sigma_{\mu\nu}\rightarrow  i\eta^a_{\mu\nu}\sigma^a$ with the $^\prime$t Hooft symbol satisfying $\eta^a_{\mu\nu}\eta^b_{\mu\nu}=4\delta^{ab}$, and  (\ref{24X})
can be simplified further 

\begin{widetext}
\bea
\label{26X}
{\cal V}^{L+R}_{qq}=\,{\kappa}_2\,A_{2N}\,
\bigg({\rm det}(UD)-4B_{2N}\,{\rm det}(U^aD^a)\bigg)
+(L\leftrightarrow R)
\eea
\end{widetext}

Another possible ``looping" of quarks is possible if we  account for nonzero quark masses, in particular the largest strange quark mass $m_s$. 
As a result, the effective $ud$ interaction (containing the $m_s$ term) is $stronger$ 
than $us,ds$ interactions, as they only contain negligibly small light quark masses. As noticed by one of us and Rosner
\cite{Shuryak:1988bf}, the ratio of the coefficients for light-light and light-strange 4-fermion terms is
\begin{equation}  { G(\bar u u \bar d d) \over  G( \bar u u \bar s s) }= {m_q^{*} +m_s \over m_q^{*} }\approx 1.5
\end{equation}
One consequence of this  is the  relatively large
violation of $SU(3)_f$ symmetry. Another is that the 
't Hooft-operator-induced  spin-dependent forces  have flavor ratios similar to the  magnetic moment ratios
(provided one ignores the difference between
the onshell ``constituent quark mass" and the determinantal masses $m^*$.) Therefore,
one cannot trace their origin simply by the mass dependence.

Now let us define $m^*$ in the Lagrangian coefficient, following \cite{Faccioli:2001ug}, see also \cite{Forkel:1994pf,Blotz:1996ad,Braguta:2003hd}. 
If there  is a single instanton in the ``empty" vacuum, as considered in 
the original 't Hooft paper, those are just quark masses $m_q$. The product of the quark masses is however
also present in the instanton density (in this setting), and they all cancel out, leaving a finite Lagrangian
even in the chiral limit $m_q\rightarrow 0$. 

In an  ensemble of instantons, the quark propagator in the QCD vacuum and restricted to the zero modes, is  approximated by  the 
form 

\begin{equation}S(x,y)=S_Z(x,y)+\sum_{I,J} \psi_{0I}^*(x) \bigg( {1 \over T} \bigg)_{I,J} \psi_{0J}(y)
\end{equation}
where $T_{IJ}$ denotes the so called ``instanton hopping" matrix,  constructed out of the Dirac
zero modes overlaps between neighboring  instantons $I,J$. Note that
it contributes as an $inverse$ matrix, as propagators are inverse to Dirac operators.
So, when one discusses a process in which
both points $x,y$ are inside the same instanton $I^*$,  as per the definition of a hard block,
we  can restrict the sum to only the term with the zero 
mode of this very instanton. This leads to the following
re-definition of the  ``determinantal mass"  

\begin{equation}\frac 1{m_u}\equiv \bigg< \bigg( {1 \over T} \bigg)_{I^*I^*}    \bigg>  \end{equation}
Furthermore, in the diagrams containing $two$ quark propagators of {\em different flavors} one has a different
averaging

\begin{equation}
\frac 1{m^2_{uudd}}\equiv \bigg< \bigg( {1 \over T} \bigg)^2_{I^*I^*}    \bigg>
 \end{equation}
These two quantities were evaluated in the random and interacting instanton liquid model, with

\begin{equation}
\frac 1{m_u^2}\approx {1 \over (177 \, {\rm MeV})^2}\ll  \frac 1{m^2_{uudd}}\approx {1\over (103\, {\rm MeV})^2}
\end{equation}

The chief consequence of these substantial  deviations from mean field can be captured by 
a ``'t Hooft operator enhancement factor"
\begin{equation}
f_{\rm tHooft}\equiv \bigg({m_{u} \over m_{uudd}}\bigg)^2 \approx 3 
    \label{eqn_hooft_inhancement} 
\end{equation}
This  enhancement  of 4-fermion operator relative to 2-fermion squared 
has also been observed on the lattice
 in~\cite{Faccioli:2003qz}.

\subsection{Spin and flavor-dependent interactions from zero-modes}

Since the t$^\prime$ Hooft induced interaction is non-local, there are additional contributions besides the local
terms retained earlier. More specifically, for two flavors its generic form  is

\begin{widetext}
\bea
\label{TNL1}
&&{\cal L}=\kappa_2 A_{2N}\,
\int \prod_{i=1}^4\frac{d^4k_i}{(2\pi)^4}\bigg(\frac{M(k_i)}{M(0)}\bigg)^{\frac 12}\,(2\pi)^4\delta^4(k_1+k_3-k_2-k_4)\nonumber\\
&&\times\frac 12\epsilon^{f_1f_2}\epsilon^{g_1g_2}
\bigg(\overline\psi_{Rf_1}(k_1)\psi_{Lg_1}(k_2)\overline\psi_{Rf_2}(k_3)\psi_{Lg_2}(k_4)
-4B_{2N}\overline\psi_{Rf_1}(k_1)\sigma^a\psi_{Lg_1}(k_2)\overline\psi_{Rf_2}(k_3)\sigma^a\psi_{Lg_2}(k_4)\bigg)
+(L\rightarrow R)\nonumber\\
\eea
with form-factors, from a Fourier transform of the quark zero mode 

\be
\label{MZEROK}
\frac{M(k)}{M(0)}=\bigg(2z\bigg[I_0(z)K_1(z)-I_1(z)K_0(z)-\frac 1z I_1(z)K_1(z)\bigg]\bigg)^2_{z=k\rho/2}
\ee
We note that  the antisymmetric operator for flavor exchange
can be rewritten as
\be
\epsilon^{f_1f_2}\epsilon^{g_1g_2}=\frac 12 \bigg(\delta^{f_1g_1}\delta^{f_2g_2}-\tau^{a, f_1g_1}\tau^{a, f_2g_2}\bigg)
\rightarrow \frac 12 (1-\tau_1\cdot \tau_2)
\ee
so that (\ref{TNL1}) reads

\bea
\label{TNL2}
&&{\cal L}=\frac 12 \kappa_2 A_{2N}\,
\int \prod_{i=1}^4\frac{d^4k_i}{(2\pi)^4}\bigg(\frac{M(k_i)}{M(0)}\bigg)^{\frac 12}\,(2\pi)^4\delta^4(k_1+k_3-k_2-k_4)\nonumber\\
&&\bigg[\bigg(\overline\psi(k_1)\psi(k_2)\overline\psi(k_3)\psi(k_4)+\overline\psi(k_1)\gamma_5\psi(k_2)\overline\psi(k_3)\gamma_5\psi(k_4)\nonumber\\
&&-\overline\psi(k_1)\tau^a\psi(k_2)\overline\psi(k_3)\tau^a\psi(k_4)-\overline\psi(k_1)\tau^a\gamma_5\psi(k_2)\overline\psi(k_3)\tau^a\gamma_5\psi(k_4)\bigg)\nonumber\\
&&-4B_{2N}
\bigg(\overline\psi(k_1)\sigma^a\psi(k_2)\overline\psi(k_3)\sigma^a\psi(k_4)+\overline\psi(k_1)\sigma^a\gamma_5\psi(k_2)\overline\psi(k_3)\sigma^a\gamma_5\psi(k_4)\nonumber\\
&&-\overline\psi(k_1)\sigma^a\tau^b\psi(k_2)\overline\psi(k_3)\sigma^a\tau^b\psi(k_4)-\overline\psi(k_1)\sigma^a\tau^b\gamma_5\psi(k_2)\overline\psi(k_3)\sigma^a\tau^b\gamma_5\psi(k_4)\bigg)
\bigg]
\eea
\end{widetext}

\begin{figure}[h!]
	\begin{center}
		\includegraphics[width=8cm]{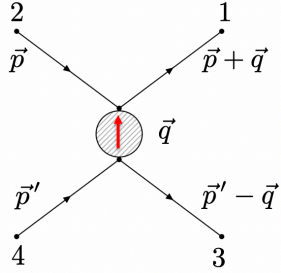}
		\caption{Kinematics in the center of mass frame of the  two-flavor instanton induced interaction with in-out on-shell Dirac fermions.}
		\label{fig_kin}
	\end{center}
\end{figure}

Going ahead of ourselves, we note that in the sequel on the light-front Hamiltonian,
we will use this operator with light cone kinematics, $ultrarelativistic$ on-shell spinor,
in spin or chiral basis. In this paper, however, we are in the center of mass frame, in
the non-relativistic setting of a constituent quark model. Therefore, we 
 perform the non-relativistic reduction of (\ref{TNL2})  with on-mass-shell  normalized Dirac spinors

\be
\label{UV}
U_s(p)=
\begin{pmatrix}
f(p)\chi_s \\
g(p)\sigma\cdot p\chi_s
\end{pmatrix}
\qquad
V_s(p)=
\begin{pmatrix}
g(p)\sigma\cdot p\chi^C_s \\
f(p)\chi^C_s
\end{pmatrix}
\ee
with standard normalization factors
\be
f(p)=\bigg(\frac{E_p+m}{2E_p}\bigg)^{\frac 12}=(E_p+m)g(p)
\ee
and $E_p=(p^2+m^2)^{\frac 12}$.  For the reduction to central, spin-orbit and spin-spin interactions, we set the kinematics in the center of mass defined in Fig.~\ref{fig_kin} with $k_1=(E_{p+q}, \vec p+\vec q)$,  $k_2=(E_p, \vec p)$, $k_3=(E_{p^\prime -q}, \vec p^{\,\prime}-\vec q)$, 
$k_4=(E_{p^\prime}, \vec p^{\,\prime})$, insert (\ref{UV}) into (\ref{TNL2})  and expand up to order $p^2/m^2$. Recall that in a meson state $\bar q q$,
the labels in Fig.~\ref{fig_kin} refers also to the flavor and spin $1,2\rightarrow U$ with mass $m_Q$ for the quark,  and $3,4\rightarrow V$ with mass $m_{\bar Q}$
for the anti-quark. More specifically after setting $m=m_Q=m_{\bar Q}$, the order $q^0/m^0$  amounts to the two-body central and spin-spin operators

\be \label{eqn_nonrel_Hooft}
-\frac 14 |\kappa_2| A_{2N}\bigg(1-\tau_1\cdot \tau_2\bigg)\bigg(1-16B_{2N}\vec S_1\cdot \vec S_2\bigg)
\ee
in the U(1) or $\eta^\prime$ channel.
For clarity, the form factor is temporarily set to 1 (zero size approximation).  Taking into account the  overall negative sign in  $\kappa_2$,
and  $16B_{2N}=4/5$, and the values of $\vec S_1\cdot \vec S_2=-3/4$ for $S=0$ states,
one finds that the last bracket simplifies to $-8/5$ for  the  pseudoscalar (pion) channel. 
Note that (\ref{TNL2}) is only active in the $(\sigma, \pi, \sigma_5, \pi_5, \eta^\prime)$ channels.
(\ref{TNL2}) does not contribute to the vector channels.

The relativisitc correction  $\sim q^2/m^2$  amounts to the two-body 
operators (with again the form factor set to 1)

\bea
&-&\frac 14 |\kappa_2| A_{2N}\bigg(1-\tau_1\cdot \tau_2\bigg)\bigg(\frac 1{4m^2}\bigg)\bigg(-L_S -4 S_T  \nonumber \\
&+&4B_{2N}\,(3L_S+4S_T+8q^2S_1\cdot S_2)\bigg)  
\eea
with spin-orbit and tensor operators
\bea
L_S&=&(S_1+S_2)\cdot (iq \times  (p-p^\prime))\\
S_T&=&S_1\cdot q S_2\cdot q, \,\,\,\, S_{12} =3 (S_1 \hat q)(S_2 \hat q)- (S_1 S_2) \nonumber
\eea
with $L_S$ symmetrized using $\vec p+\vec p^{\,\prime}=0$.
The translation $iq\rightarrow \nabla$ to configuration space will be understood.

\subsection{Zero-mode-induced potential for pions}

In section \ref{sec_fixing_G}  we have already studied the pion case, and defined
the value of the 't Hooft effective coupling (\ref{eqn_G_pion}) needed to put the pion at the empirical
mass.  The first question we now discuss is whether the ILM (component of the instanton ensemble) can
provide a coupling of this magnitude. 
The answer to this question is affirmative.

The ``determinantal masses" induced by the quark condensates are given by the following expressions for light and strange quarks
\begin{equation}
	m^*_q={2\pi^2\rho^2 \over 3}| \langle \bar q q \rangle |, \,\,\,\, 	m^*_s={2\pi^2\rho^2 \over 3}| \langle \bar s s \rangle|+m_s 
\end{equation}
Using them, for a fixed instanton radius $\rho=0.33 \, {\rm fm}\approx 1.4 \, {\rm GeV}^{-1}$, we get   the following value of the coupling of $\bar u d \bar d u$ 't Hooft vertex

\begin{equation}
G_{ud}=A_{2N}{3 n\over 8}{(4\pi^2)^3 \rho^9 \over (	m^*_q \rho )^2 (m^*_s \rho) } | \langle \bar s s\rangle |\approx 27 \, {\rm GeV}^{-2}
\end{equation}
It is larger than the value of $17 \, {\rm GeV}^{-2}$  we fitted
from the pion mass using the  Schroedinger equation above,  by about 50\%. 
However, note  that it contains a large (sixth) power of the instanton size, which is not yet defined with sufficiently high precision.
The numbers would match if one reduces it from $1/3\, fm$ to $ 0.29\,  fm$. 

Now, assuming that the 't Hooft-induced potential puts the pion  at the right mass, we now turn our attention to the $\rho$ meson.
In the previous section we have found that there is an attractive  't Hooft-induced potential which, at the leading
nonrelativistic order (\ref{eqn_nonrel_Hooft}), is just a {\it half} of that for the pion. Naively, one might
think that this potential will pull the mass
of the $\rho$  to {\it half} of what it does to the pion, putting it right at the experimentally observed value.
Unfortunately, it is not the case. 
The  effects of the  't Hooft operator are too large to be treated as small perturbations. A factor two difference in a potential,  does not imply a factor two in the
energy, because the wave function itself depends strongly on it. What we found is that, due to the  't Hooft-induced potential, the pion's size get reduced, 
making it comparable to the
instanton potential range $\sim \rho$. Half of it, acting against the repulsive $V_{SS(r)}$ from the earlier contributions, 
 fail to shrink the $\rho$ size, which remains roughly twice the size of $\pi$. This reduces the wave function
 at small distances roughly by a factor of $2^3$, and effectively reduces the matrix element of the $V^{\rm Hooft}(r)$ in
 the $\rho$ case by a comparable factor. The corresponding potentials and wave functions are shown in Fig.\ref{fig_rho_pi}.

\subsection{The $\rho$ meson and the ``molecule-induced" potential} \label{sec_rho}

This leaves us with the model prediction for the $\rho$ meson mass  $m_\rho$ in the vicinity of the mass following from a   ``generic Cornell meson",
with $m\approx 1.4\,\, GeV$, far from its experimental mass. The theory part is missing
something very important for the vector light quark channel. 

Fortunately, there is still one more effect, already discussed in the  literature, the second-order 't Hooft effective 
forces due to ``molecular" configurations, see the lower plot in the Introduction, Fig.\ref{fig:pi_rho} . In particular, the effective Lagrangian
induced by these molecules  was used in   the theory of color superconductivity at high density \cite{Rapp:1999qa}, from which
we borrow its analytic form

\begin{widetext}
\bea
  L_{\rm mol}/G_{\rm mol} =&&
{1\over N_c^2} \big[ (\bar \psi \gamma_\mu \psi)^2+(\bar \psi \gamma_\mu \gamma_5\psi)^2 \big]  
-{1 \over 2N_c(N_c-1)}\big[ (\bar \psi \gamma_\mu \lambda^a \psi)^2+(\bar \psi \gamma_\mu \gamma_5\lambda^a\psi)^2 \big] \nonumber\\
&&-{1 \over N_c^2}\big[ (\bar \psi \gamma_\mu \psi)^2-(\bar \psi \gamma_\mu \gamma_5\psi)^2 \big]  
-{2 N_c-1 \over 2N_c (N_c^2-1)}\big[ (\bar \psi \gamma_\mu \lambda^a \psi)^2-(\bar \psi \gamma_\mu \gamma_5\lambda^a\psi)^2 \big] 
\eea
\end{widetext}
It follows by using the same  color orientation for the instanton and anti-instanton, which locks them to their most attractive channel.
The quark fields come from zero modes of $I$ and $\bar I$, so there is no Pauli principle and the flavors can be both the same or
different. Note that in in this case the chiralities of the quarks and
antiquarks are the same in each term, which is suitable for the $\rho$ meson channel.
Choosing the vector $\gamma_\mu \lambda^a$ terms and reducing the color matrices for the meson case,  we can reduce it to

\begin{equation} 
\label{MOLX}
L_{\rm mol}=G_{\rm mol} (\bar u \gamma_\mu u) _x (\bar d \gamma_\mu d)_y F_{mol}(x-y) 
\end{equation} 
where we introduced another form-factor to account for the  nonlocality of  the  instanton-antiinstanton molecule.
In principle, one can evaluate (\ref{MOLX})  in a way that is analogous to our treatment of 
the  't Hooft Lagrangian. However, it would  now include  $two$ 4-dimensional integrations over positions of
both centers, plus the relativiely complicated $I\bar I$ interaction depending on the relative color orientations. Qualitatively, one expects its non-locality size
to approximately double, so we may just  substitute the instanton size $\rho \rightarrow 2\rho$ in the instanton form factor as an estimate.
As a result,  its coefficient is reduced by $1/2^3$ in the formfactor (\ref{eqn_FF}).  Such a  size reduction 
of the potential range is precisely what is needed to obtain the correct $\rho$ meson mass.

More specifically, setting  the corresponding potential into the Schroedinger equation, along with the Cornell and spin-spin terms, 
we  find that the correct $\rho$ mass is indeed obtained for

 \begin{equation}
{G_{\rm mol} \over G_{\rm Hooft}}= {120\, {\rm GeV}^2 \over 17\, {\rm GeV}^2} \sim { n_{\rm mol} +n_{\rm ILM} \over n_{\rm ILM}}\approx 7
\end{equation}
Note that this  ratio of couplings (or densities) agrees well with what was 
  deduced from the central potential fitting the linear potential discussed above. 

(As a parting comment, we remind the reader that the NJL model -- the pioneering effort to
describe chiral symmetry breaking and nonperturbative interactions in hadrons --
involves flavor-diagonal terms like the one we used here. In later papers 
based on the NJL  model,  those contributions were attributed to the  ``strong coupling part" of one  gluon exchange.)

\begin{figure}[b!]
\begin{center}
\includegraphics[width=6cm]{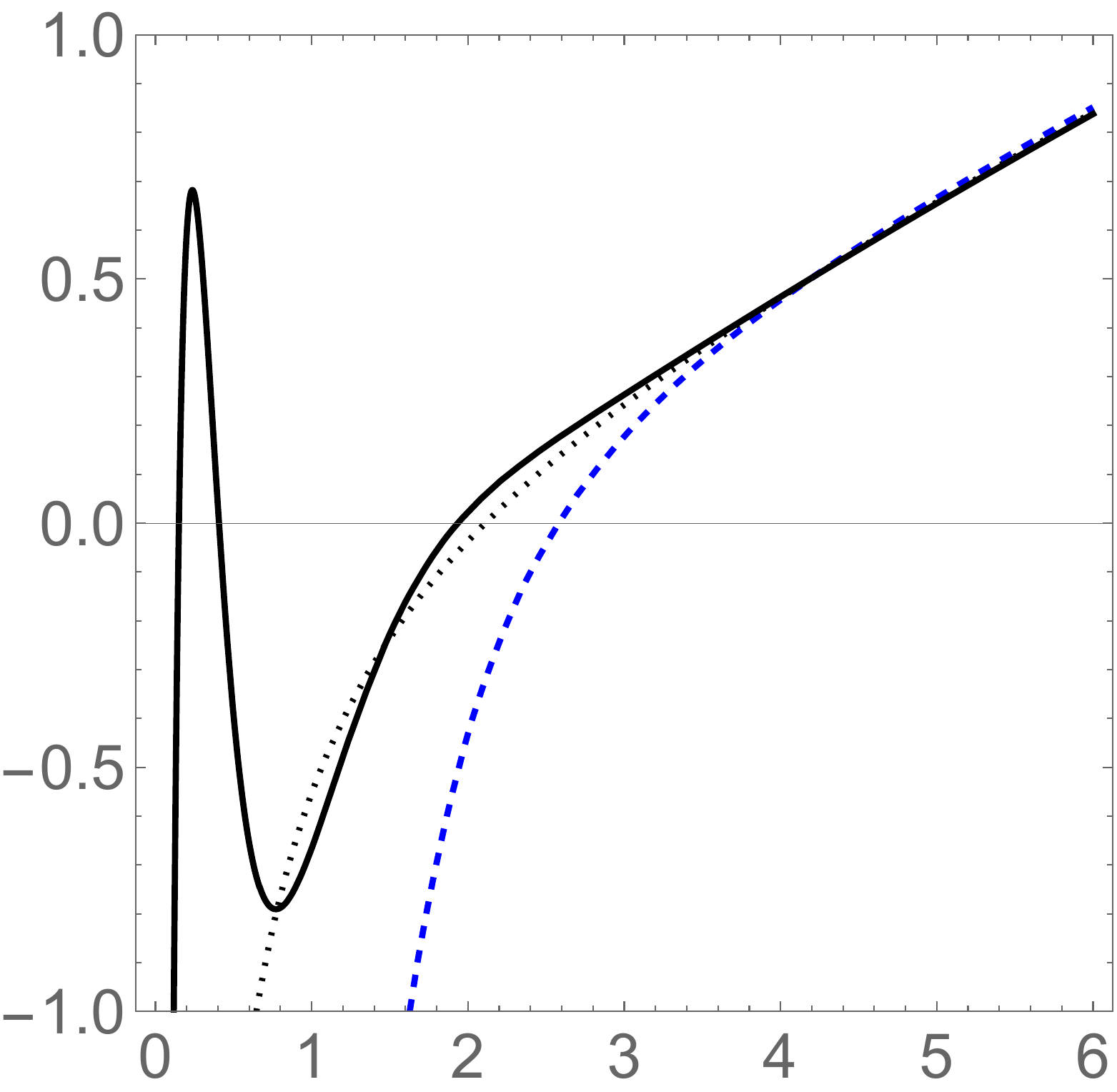}
\includegraphics[width=6cm]{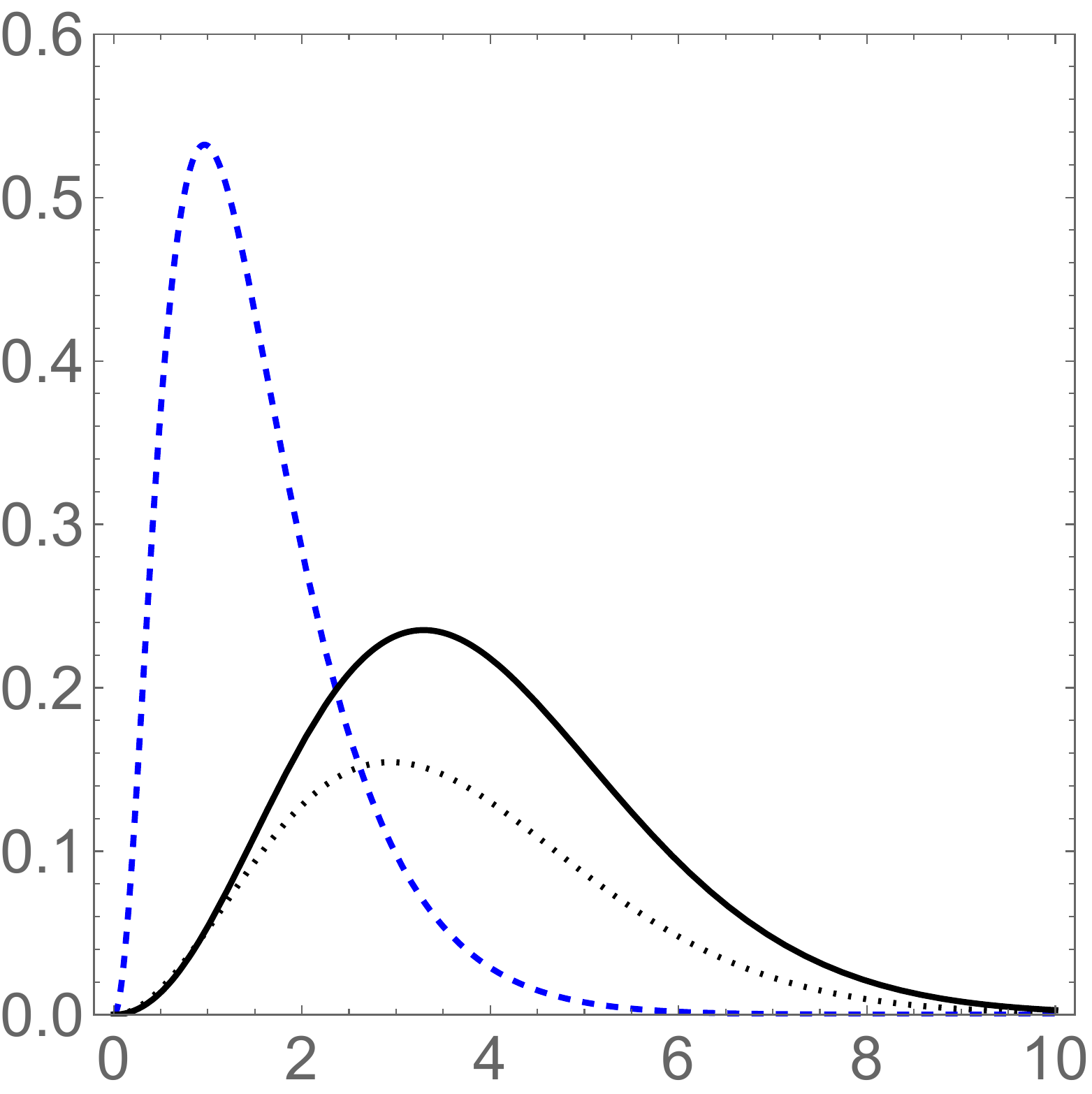}
\caption{The total potentials  including the 't Hooft ones (upper plot) and wave functions as $r^2\psi(r)^2$ (lower plot) for the $\rho$ channel
(solid curves), the $\pi$ channel (blue dashed curves) and the generic Cornell potential (black dots).
All potentials  are shown versus the distance $r$ in ${\rm GeV}^{-1}$.
}
\label{fig_rho_pi}
\end{center}
\end{figure}

\subsection{Zero mode effects in heavy-light systems} \label{sec_heavy_light}

The spin interactions for heavy-light quarks split into the contributions solely from the non-zero modes which give rise to the same
general spin dependent interactions as in (\ref{SD}).

For simplicity, we start with the heavy-light mesons, with a single light quark. 
 The contributions  to the quark propagators come 
 from the zero-mode for the light quark, 
and the non-zero-modes for the heavy-quark. The resulting spin-independent interaction 
was derived in \cite{Chernyshev:1995gj}

\begin{widetext}
\bea
\label{QLIGHT1}
{\cal L}_{qQ}=-\bigg(\frac{\Delta m_Q\Delta m_q}{2nN_c}\bigg)
\bigg(\overline {\bf Q}\frac {1+\gamma^0}2{\bf Q}\, \overline {\bf q} {\bf q} 
+\frac 14\overline{Q}\frac {1+\gamma^0}2\lambda^a {\bf Q}\,\overline {\bf q}\lambda^a {\bf q}\bigg)
\eea
while the spin-dependent interaction  is more suppressed

\bea
\label{QLIGHT2}
{\cal L}^{\rm spin}_{qQ}=-\bigg(\frac{\Delta m^{\rm spin}_Q\Delta m_q}{2nN_c}\bigg) 
\times
\frac 14 \bigg(\overline {\bf Q}\frac {1+\gamma^0}2\lambda^a\sigma^{\mu\nu}{\bf Q}\, 
\overline {\bf q} \lambda^a\sigma^{\mu\nu}{\bf q}\bigg) \nonumber
\eea
\end{widetext}
with $\Delta m^{\rm spin}_Q/\Delta m_q\sim 1/100$ (charm).
The local potential stemming from (\ref{QLIGHT1}-\ref{QLIGHT2}) is 

\bea
V_{qQ}(r)=\bigg(\frac{\Delta m_Q\Delta m_q}{2nN_c}\bigg)\bigg(1+\frac 14\lambda_q^a\lambda_Q^a\bigg)\,\delta^3(r)
\eea
In the mesons, the colors on the two lines are opposite and the second term dominates the first, so this local potential is negative.
For  the $D$ meson it amounts to  $\langle V_{qQ} \rangle\approx -180 \, {\rm MeV}$ \cite{Chernyshev:1995gj},
to be compared with the non-relativistic energy due to the standard Hamiltonian, kinetic plus
central potential average $$\langle P^2/2m_q+V_c (r)\rangle\approx +250 \, {\rm MeV}.$$ 

The spin-spin potential is

\bea
V^{\rm spin}_{qQ}(r)= -\bigg(\frac{\Delta m^{\rm spin}_Q\Delta m_q}{2nN_C}\bigg)\,{ S}_q\cdot { S}_Q\,\lambda^a_q\lambda_Q^a\,\delta^3(r)\nonumber\\
\eea
For the D meson both the relative spin and color are negative, so the contribution of this term  is negative
as well, giving $\langle V^{\rm spin}_{qQ} \rangle\approx - 42 \, {\rm MeV}$ \cite{Chernyshev:1995gj}.
The contribution of this term to the spin splitting is then
$$M(D^*)-M(D)=(4/3)42 \, {\rm MeV}=56 \, {\rm MeV}$$
to be compared with  our calculations above 
giving together $\langle V^{\rm pert}_{SS}+ V^{\rm inst}_{SS}\rangle\approx  96\, {\rm MeV}$. 
We thus conclude that all the three contributions together  -- the perturbative one and the two instanton-induced ones -- do finally reproduce this mass difference,  close to the experimental value $M(D^*)-M(D)=137\, {\rm MeV}$.

\section{Summary and Discussion} \label{sec_summary}

The purpose of this work was to have a fresh look at  the origins of the central and spin-dependent potentials,
used in the non-relativistic description of mesons. We focused  in the range of  ``intermediate distances" $r\sim 0.2-0.5\, {\rm fm}$,
where the  spin-dependent forces are mostly active.

The phenomenological part of this paper focused solely on  mesons, from heavy charmonia to 
heavy-light and to light-light ones. Using data for the spin splittings in the 1S shell (2 states) and the 1P shell (4 states), 
we evaluated matrix elements of the individual  spin-dependent potentials. Among the important observations is a change of
 sign of the tensor force for lighter systems. 

The theory part started with a reformulation of the instanton ensemble. Unlike previous studies 
using the  dilute ILM, and mostly focused on the spontaneous breaking of  chiral symmetry  and zero modes, 
the ensemble we consider now includes $I \bar I$ "molecules". 
It was shown that in this regime the instanton-induced potential reproduces the magnitude of the central potential, 
up to distances $r < 1\, {\rm fm}$. Ascribing the central potential $V_C(r)$  at $r < 1\, {\rm fm}$ to this extra contribution, we get its
normalization fixed. It turns out to be about twice less dense than the lattice extrapolations of Fig.~\ref{fig_cooling}(b), 
with a diluteness parameter $\kappa\sim 1$. 
 
This leads to evaluations of the nonperturbative part of spin-dependent 
potentials.  Their application to heavy-heavy systems can be considered consistent with available
lattice data for $V_{SS}(r)$, and also with phenomenology. 
 One important
finding  is that the perturbative and  instanton-induced contributions are {\em comparable in magnitude}.
Together they semi-quantitavily reproduce lattice-based potential and the spin splitting in heavy quarkonia.

Going further from heavy to light systems, we analyzed four P-shell states, attempting
to extract spin-orbit and tensor potentials. The reader should however be aware that for light quarks,
we no longer are well below the open decay channels, whose possible effects on observed masses 
have  been ignored for now. Experience with some mesons in the past  suggests that these effects are 
small. An argument that we mostly have it correct is that, as one can see from Table \ref{tab_1P},
that for all three potentials the trend, from heavy to light quarks, seems to be consistent.

Although our straightforward evaluation of perturbative and 
instanton-induced potentials have not reproduced the apparently negative tensor force 
for light quarks,   the negative $sign$ of the instanton-induced term indicates that, with
some play of parameters, it perhaps can be done. 

Furthermore, for light quarks there exists  ``anomalous" processes,
in which quarks do not travel continuously in time, but are instead dumped into the Dirac sea and
substituted by another quark, with different value of the chirality. The non-relativistic reduction of 6-quark 't Hooft
operator  have produced novel spin-dependent forces. Their
application to the pion case is successful, as well as to the heavy-light systems. 
The effects related to the zero modes and the  't Hooft effective interactions can still be put in the form of quasi-local flavor-nondiagonal
potentials, i.e. central and spin-dependent. The
phenomenological
implications of this is a reminder of the  {\em two component} instanton model discussed here, 
meaning that for the 't Hooft potential the instanton  density is the smaller one, taken from the original ILM. 
The magnitude of such  't Hooft potential is found of the size needed to put the pion mass at its small physical value.
It is in agreement with the long-known view of the pion as depicted in the upper Fig.~\ref{fig:pi_rho}.

The case of the $\rho$ meson is more complicated. Even with the   't Hooft potential  its mass remains
well above  the observed value. However, there is 
the so called  ``molecular" component, schematically shown in the lower part of Fig.~\ref{fig:pi_rho}.
It generates attractive forces diagonal in chirality. Assuming that with the molecular component the 
density is 7 times larger than that of the original  ILM, we get an estimate of its coupling which we find to 
put the $\rho$ meson in  the correct mass range.

The focus of this work  has not been on  numerical accuracy,  but rather  on the  microscopic  origin of the central and spin-dependent forces. It should be viewed as an introduction to the use of  these concepts for the light front Hamiltonians and wave functions, which we will address next.

 In sum, the paper proposes novel interpretations of interquark potentials, that
 should and will be tested. One way  to do this is to 
calculate all spin-dependent potentials on the lattice {\it during}
a  ``cooling" process (e.g. via gradient flow), and assess how the results
depend on the cooling time (resolution). This will allow for a quantitative discrimination between  the perturbative (gluon-induced),
 and topological (instanton-induced) effects, at low resolution.

\section{Acknowledgements}
This work was supported by the U.S. Department of Energy under Contract No.DE-FG-88ER40388.

\vskip 1cm
\appendix

\section{Asymptotics of $V_C(r)$}

The instanton induced central potental $V_C(r)$ in Fig.~\ref{fig_two_potentials} asymptotes 
twice the self-energy $2\Delta m_Q$.
This asymptotics is reached from {\it above} in singular gauge contrary to expectations. 
The simplest way to see  this,  is  to note that

\begin{widetext}
\begin{equation}
2\Delta M_Q \rho= {8\pi n \rho^4 \over N_c}\int_0^\infty dy y^2\bigg(1+cos\bigg({\pi y \over \rho^2+y^2} \bigg) \bigg) 
\approx {8\pi n \rho^4 \over N_c}  2.2
\end{equation}
\end{widetext}
 is substantially smaller than the potential in Fig.~\ref{fig_two_potentials}.
A numerical  evaluation of $I(x)$ as in (\ref{eqn_I_of_x}) at large $x$,  is  shown in Fig.~\ref{fig_I_of_r}.
It confirms that the asymptotics is slowly reached from above. 

\begin{figure}[htbp]
\begin{center}
\includegraphics[width=7cm]{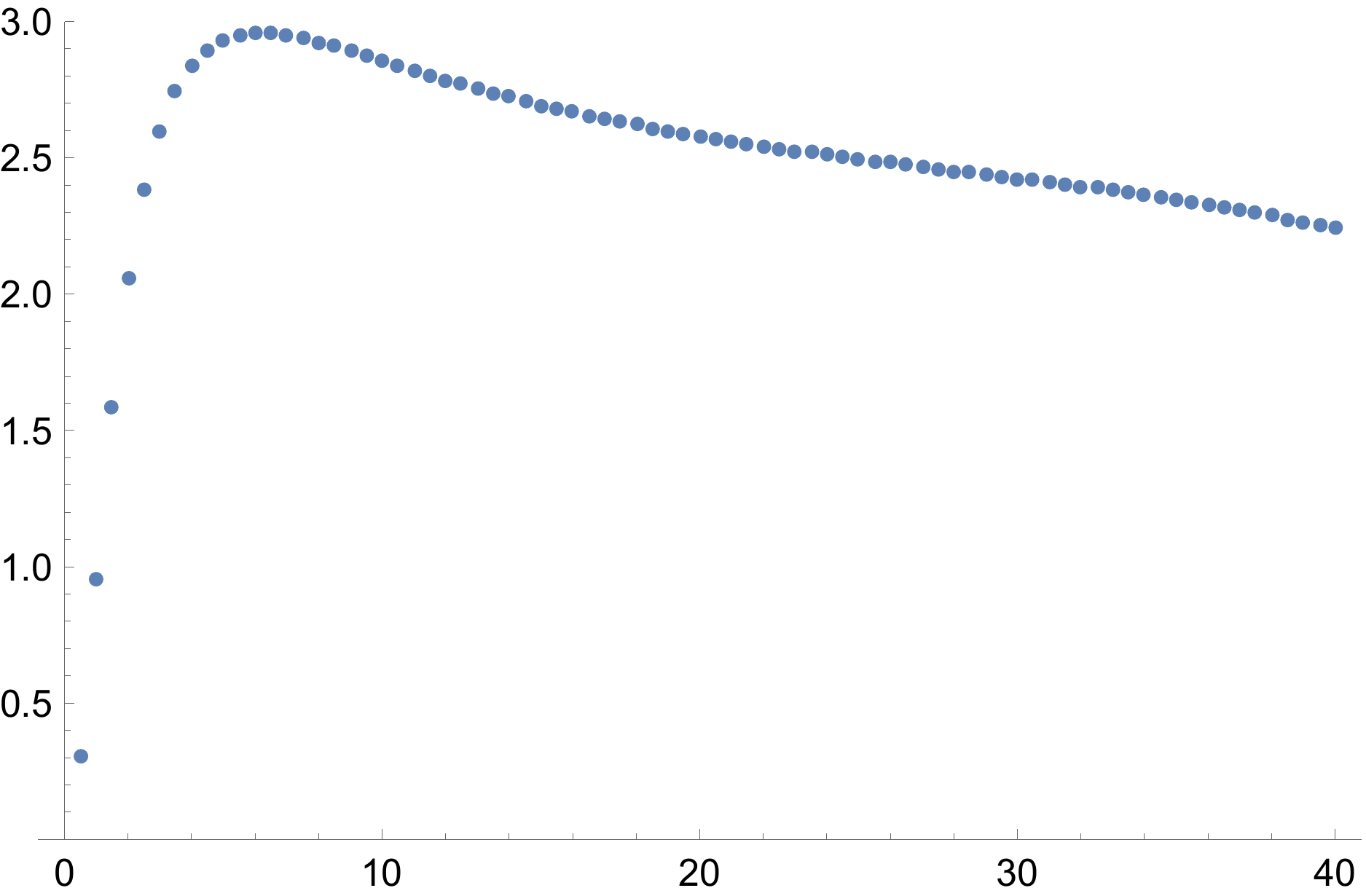}
\caption{$I(x)$ versus $x$ as given in (\ref{fig_two_potentials})}
\label{fig_I_of_r}
\end{center}
\end{figure}

We note that a Taylor expansion of the integrand  in $I(x)$ suggests $1/r^3$ as the leading contribution,
but the remaining integration is divergent, thereby invalidating the expansion. In \cite{Yakhshiev:2016keg} 
it was suggested that the asymptotics is reached from below through $-\pi^2/2r$. This result is correct in
a perturbative regular gauge with no additional color rotations, 
but not in a non-perturbative singular gauge with additional color rotations.  Also, the $1/r$ central potential yields a spin-spin potential $V_{SS}(r)$ as a delta-function
which is also not compatible with the $1/r^2$ behavior noted above. Fig.~\ref{fig_I_of_r} can be fitted
numerically by $2\Delta M_Q+ C/r^p$ with  $p\ll 1$ and $C>0$. Its ensuing  Laplacian then gives
$p(1-p)/r^{2+p}$, which is consistent with $V_{SS}(r)$ at large $r$.

\section{Definitions of  the spin-dependent potentials} \label{sec_correlators}

Following Ref.~\cite{Eichten:1980mw}, we give for completeness the correlation functions
generating the five spin-dependent potentials. The integrals over points $z,z'$ are taken along two
Wilson lines located at $\vec x_1,\vec x_2$  and running 
 in the Euclidean time direction,   from $-T/2$ to $T/2$. 
 The limit of large $T\rightarrow \infty $ is not written explicitly but assumed. The normalization to $\langle 1 \rangle$ means subtraction
of the central potential coming from the correlator of Wilson lines without the extra field strengths

\begin{widetext}
\bea
\label{XXX}
 {r^k \over r} {dV_1 (r)\over dr}=&&\epsilon_{ijk} \int dz dz' ({z-z' \over T}){g^2 \over 2}
\langle B^i(\vec x_1,z) E^j(\vec x_1,z')\rangle /\langle 1 \rangle\nonumber\\
 {r^k \over r} {dV_2 (r)\over dr}=&&\epsilon_{ijk} \int dz dz' ({z-z' \over T}){g^2 \over 2}
\langle B^i(\vec x_2,z) E^j(\vec x_1,z')\rangle /\langle 1 \rangle\nonumber\\
&&\times \big[ (\hat r^i \hat r^j -\delta_{ij}/3)V_3(r)+\delta_{ij}/3)V_4(r)\big]
=\int dz dz' 
{g^2 \over T}\langle B^i(\vec x_2,z) E^j(\vec x_1,z')\rangle /\langle 1 \rangle 
\eea
\end{widetext}
Although it  involves only the  vector product of electric and magnetic fields, using the Bianchi identity one can 
express it via a correlator of two $B$ fields, as expected for non-relativistic spin interactions.
It is however of little importance for instantons, which are (anti) self-dual $\vec E =\pm \vec B$.

Note that the reduction of the dressed Wilson line ``WEW" line with an electric field insertion can be made into a derivative of the Wilson line
over its location, via the following steps

\begin{widetext}
\bea
\label{ID2}
&&D_{m}(y){\bf W}(x_0,y_0)-{\bf W}(x_0,y_0)D_{m }(y)\nonumber\\
&&=\langle x_0 | D_{0 }\frac 1{iD_0}-\frac 1{iD_0}D_{m}|y_0\rangle \nonumber\\
&&=
\langle x_0|\frac 1{iD_0} [iD_0, D_{m }]\frac 1{iD_0}|y_0\rangle\nonumber\\
&&=\langle x_0|\frac 1{iD_0} (- F_{0 m})\frac 1{iD_0}|y_0\rangle=\int_{-\frac 12 T}^{+\frac 12 T} dz_0\,{\bf W}(-T,z_0)(- F_{0 m})(z_0){\bf W}(z_0,T)
\eea
\end{widetext}
and in the large $|T|\rightarrow \infty$ asymptotic the feld is assumed to vanish, so the covariant derivative can be changed to ordinary one.

Subsequent discussions of the ${\cal O}(1/m^2)$ of the potentials have been made,
starting from complete
effective actions of NRQCD  and  pNRQCD to this order, and some  modifications
of the expressions  in~\cite{Eichten:1980mw} were found (See Eqs.~49-51 in
\cite{Pineda:2000sz}). Modulo matching, the only difference was noted in the
spin-orbit coefficient with no $\frac 12$ in (\ref{XXX}).  However, for the purposes of our work,  it is enough to note that the
spin-spin and spin-orbit potentials are related to the correlators of two magnetic fields, and the
spin-orbit to the correlator of electric and magnetic fields. (Of course, the correlators include the
appropriate Wilson lines, and are integrated over  the time difference between the field strengths.)

\begin{figure}[htbp]
\begin{center}
\includegraphics[width=6cm]{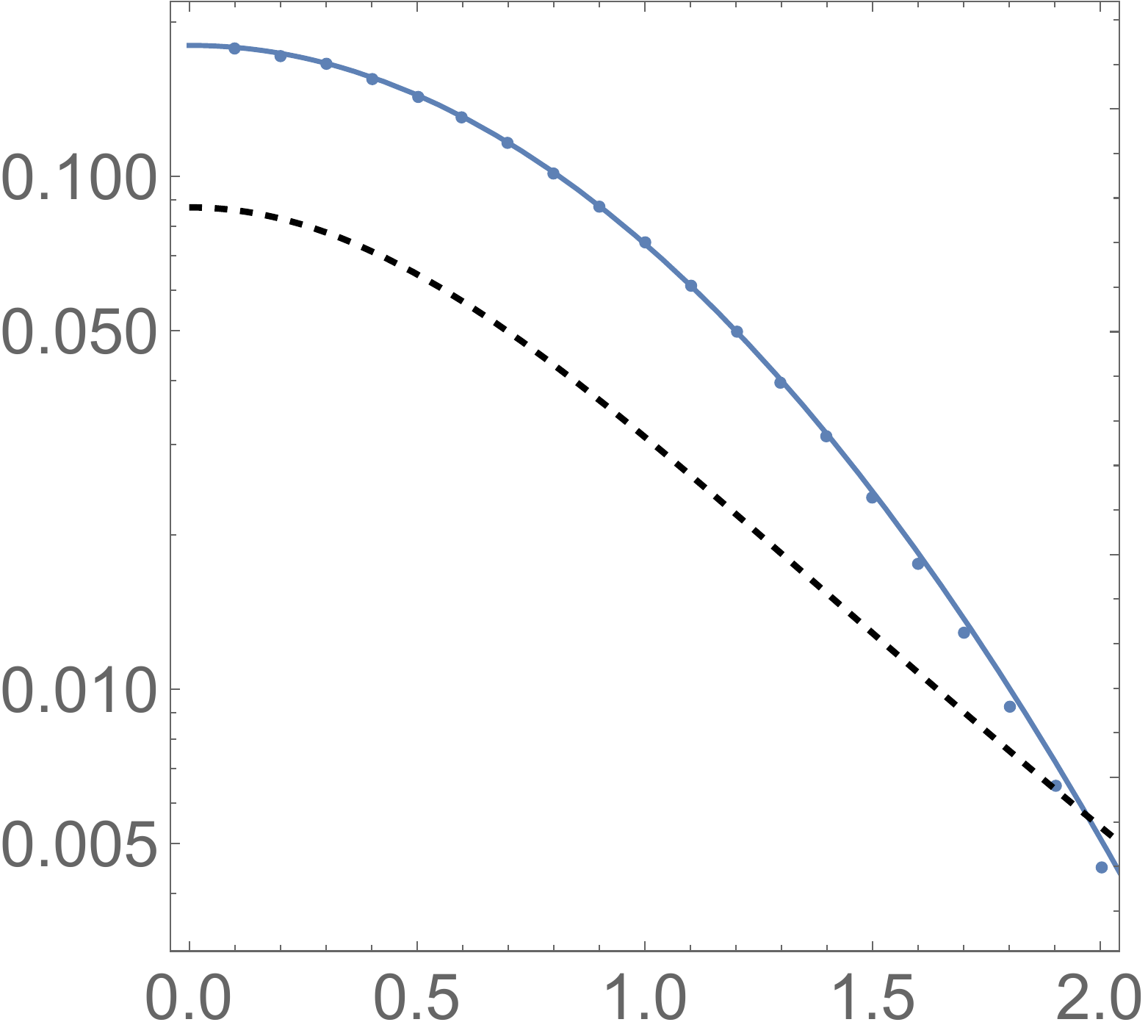}
\caption{Numerically integrated and normalized integral (\ref{eqn_FF}) (points), compared to a  fit (solid line). Also for  comparison we show (the dashed line),  the normalized ``smoothed delta function" as used in (\ref{regulated_Laplacian}) for $\delta=\rho$. It was our initial choice, but as shown here, it does not provide an accurate description of the form-factor and  therefore is not used. }
\label{fig_F_of_r}
\end{center}
\end{figure}

\section{Form-factor in the 't Hooft Lagrangian} \label{sec_nonlocality_of_Hooft}

In many applications, the instanton induced form-factor on the tunneling quarks, is used in the momentum representation.
More specifically, as the Fourier transform of the zero modes in (\ref{MZEROK}). 
However, in  the effective potential we use, there are four zero modes, thus the fourth power of $M(k)$. Its inverse 
Fourier transform to coordinate space is complicated. We use instead directly the coordinate expression,
with two  $densities$ of zero modes separated by a distance $r$ , with the explicit integration over the location 
of the instanton center $z_\mu$

\begin{widetext}
 \bea
 \label{eqn_FF}
 F(r)\sim&&  \int d^4z  |\psi_0(z^\mu -r^\mu/2)|^2  |\psi_0(z^\mu +r^\mu/2)|^2 \nonumber \\
 \sim &&
 \int {z^3 dz sin^2(\theta) d\theta   \over (z^2+r^2/4+z r cos(\theta)+\rho^2)^3  (z^2+r^2/4-z r cos(\theta)+\rho^2)^3} 
 \approx  0.18\,{\rm exp}\big(-0.89\, r^2 \big)
\eea
The  last expression is a fit, normalized by $\int d^3 F(r)=1 $, with our standard size $\rho=1.4\, {\rm GeV}^{-1}$.
The comparison between the numerical integral and the fit is shown  in Fig.~\ref{fig_F_of_r}.
\end{widetext}

\section{Field strengths through conformal mapping} 

The field strengths  $F_{MN}$
and its dual $\star {F}_{MN}$  for the tunneling at fixed energy in O(4) symmetric form (Step 1) are given
by

\begin{eqnarray}
\label{Z10}
y^2\vec{E}=&&\vec{\sigma}\left(f^{\prime}-\frac{(\vec{y})^2}{y^2}(f^{\prime}-2f\bar f)\right)\nonumber\nonumber \\
&&+\frac{(\vec{y}\cdot\vec{\sigma}\vec{y}-y_4\vec{\sigma} \times \vec{y})}{y^2}(f^{\prime}-2f\bar f)\nonumber\\
y^2\vec{B}=&&\vec{\sigma}\left(2f\bar f+\frac{(\vec{y})^2}{y^2}(f^{\prime}-2f\bar f)\right)\nonumber \\
&&-\frac{(\vec{y}\cdot\vec{\sigma}\vec{y}-y_4\vec{\sigma} \times \vec{y})}{y^2}(f^{\prime}-2f\bar f)
\end{eqnarray}
with $E^{i}=F^{i4}$ and $B^{i}=\frac 12{\epsilon^{ijk}}F^{jk}$. For self-dual fields
$f^\prime=2f\bar f$  and $\vec E=\vec B$ are hedgehogs in color-spin as expected for the instanton path. 
For the sphaleron path,  we have $f=\frac 12$ and $(\vec E+\vec B)= \vec\sigma/2y^2$, as only the electric plus magnetic field
sum is hedgehog in color-spin.
We then use the conformal (stereographic) map (Step 2), as explained in the text.

\bibliography{mesons_at_CM1,mesons_at_CM}
\end{document}